\documentclass[%
 reprint,
 amsmath,amssymb,
 aps,
]{revtex4-2}

\usepackage{graphicx}
\usepackage{dcolumn}
\usepackage{bm}
\usepackage[mathlines]{lineno}

\usepackage{siunitx}
\usepackage{nicefrac}
\usepackage{booktabs}
\usepackage[caption=false]{subfig}
\UseRawInputEncoding

\begin{document}

\preprint{Physical Review Accelerators and Beams}

\title{Design of a large energy acceptance beamline using Fixed Field Accelerator optics}

\author{A. F. Steinberg}
 \altaffiliation[Also at ]{The University of Melbourne, Australia} \email{adam.steinberg@manchester.ac.uk}
\author{R. B. Appleby}
\affiliation{The University of Manchester, Manchester, UK}
\affiliation{Cockcroft Institute, Warrington, UK}

\author{J. S.L. Yap}
\author{S. L. Sheehy}\altaffiliation[Also at ]{The Australian Nuclear Science and Technology Organisation (ANSTO)}
\affiliation{The University of Melbourne, Parkville, Australia}

\date{\today}

\begin{abstract}
Large energy acceptance arcs have been proposed for applications such as cancer therapy, muon accelerators, and recirculating linacs. The efficacy of charged particle therapy can be improved by reducing the energy layer switching time, however this is currently limited by the small momentum acceptance of the beam delivery system ($<$\SI{\pm 1}{\percent}). A `closed-dispersion arc' with a large momentum acceptance has the potential to remove this bottleneck, however such a beamline has not yet been constructed. We have developed a design methodology for large momentum acceptance arcs with Fixed Field Accelerator optics, applying it to a demonstrator beam delivery system for protons at \num{0.5}--\SI{3.0}{\mega\electronvolt} (\SI{\pm 42}{\percent} momentum acceptance) as part of the TURBO project at the University of Melbourne. Using realistic magnetic fields, a beamline has been designed with zero dispersion at either end. An algorithm has been devised for the construction of permanent magnet Halbach arrays for this beamline with multipole error below one part in $10^4$, using commercially available magnets. The sensitivity to errors has been investigated, finding that the delivered beam is robust in realistic conditions. This study demonstrates that a closed-dispersion arc with fixed fields can achieve a large momentum acceptance, and we outline future work required to develop these ideas into a complete proof-of-principle beam delivery system that can be scaled up for a medical facility.

\end{abstract}

\maketitle

\section{Motivation}

Over the course of the past thirty years, charged particle therapy has become a mature treatment modality, with over \num{100} proton therapy facilities worldwide and \num{300000} patients treated up to \num{2022} \cite{noauthor_particle_nodate}. Charged particle therapy is advantageous in many cases compared to conventional cancer treatment with X-rays \cite{olsen_proton_2007, burnet_estimating_2022, mohan_review_2022}: whereas an X-ray beam deposits dose following an approximately exponential fall-off, protons (usually between \num{70}--\SI{250}{\mega\electronvolt}) deliver a small entrance dose that rises sharply at a point, then rapidly drops to zero \cite{mohan_proton_2017}. The depth of this characteristic `Bragg peak' increases as a function of the beam energy. In practice, the particle beam energy is varied to create a `spread-out Bragg peak' covering the treatment volume, giving many discrete `energy layers' \cite{paganetti_proton_2016}. 

Realising the full potential of particle therapy has taken time, with most improvements arising from technological advances. One example of a significant change to enhance tumour conformality was the move away from using passively scattered beams to cover the tumour volume to pencil-beam scanning, enabling intensity modulated proton therapy \cite{mohan_proton_2017}. In addition, the footprint of proton therapy facilities has decreased, from the dual synchrotron rings with \SI{130}{\meter} circumference at HIMAC \cite{yamada_commissioning_1995} down to recent proposals for cyclotrons enabling single-room systems. In this time, treatment delivery methods have also varied, from fixed beamlines to gantries which rotate around a stationary patient to ensure the scanned beam can maximise dose to the treatment volume while minimising it elsewhere.

Recent studies have suggested that proton therapy can be made more efficient and effective by reducing the time taken to switch beam energies \cite{suzuki_quantitative_2011, shen_technical_2017, paganetti_roadmap_2021, yap_future_2021}.  This would reduce the impact of interplay effects, where patient motion relative to the beam during treatment requires mitigation by methods such as treatment gating while the patient holds their breath, beam rescanning, and increasing treatment margins to improve dose uniformity \cite{grassberger_motion_2013, engelsman_physics_2013, bert_motion_2011}. By lowering the energy layer switching time, the need for dose repainting may be reduced \cite{mohan_empowering_2017}, however this has not been well-studied experimentally as the minimum time to switch energy layers is limited by the beam delivery system. In addition, rapid energy switching would shorten treatment times, improve the patient experience, and increase patient throughput, though overall treatment times would still be dominated by patient setup \cite{suzuki_quantitative_2011}. A potential future improvement to charged particle therapy may be realised if the beam delivery time can be reduced even further, such that a dose $\gtrsim$\SI{40}{\gray} is delivered in $\lesssim$\SI{100}{\milli\second}. Studies suggest that the `FLASH effect' in this regime may enable dose escalation and hypofractionation without further damage to healthy tissue \cite{diffenderfer_current_2022, jolly_technical_2020}, however current beam delivery systems have insufficient energy acceptance to enable active beam scanning with energy layers at FLASH dose rates. In proton therapy, this has lead to `shoot through' or transmission beam studies which lose the conformality provided by the Bragg peak \cite{verhaegen_considerations_2021}. As such, enabling rapid beam delivery by minimising the energy layer switching time is a clear goal toward improved particle therapy systems in future, however new technologies will be required.

Current beam delivery systems typically have a momentum acceptance \SI{\leq 1}{\percent}, \cite{badano_proton-ion_2000} limited by their linear magnetic fields and small beampipe aperture: to vary the beam energy, all the magnets in the beamline must have their fields ramped synchronously as the beam momentum changes, which gives rise to the energy layer switching time. An ideal beamline would be able to deliver any clinical proton beam energy without adjustment: for the case of \num{70}--\SI{250}{\mega\electronvolt} protons, this requires a momentum acceptance greater than \SI{\pm 32.8}{\percent}.  Such a beam delivery system would enable energy variation as rapid as the proton source can provide. If the momentum acceptance can be increased to \SI{\pm 40}{\percent} (giving a maximum energy of \SI{330}{\mega\electronvolt}), the beamline would also be suitable for proton CT \cite{schulte_conceptual_2004}. All energies must enter and exit the beam delivery system at the same point in the transverse plane to ensure that no magnet adjustments are required: the beamline must comprise a `closed-dispersion arc'.

A closed-dispersion arc with a very large energy acceptance would have many applications beyond medicine, in cases where beam transport over a large momentum range without varying the magnet settings would be useful. One example is muon accelerators, where rapid acceleration is vital to prevent beam losses due to decay: without dispersion-free regions, the transverse size of the RF may prove prohibitively large. Another application is in recirculating linacs, where the energy gain per turn is sufficiently large that the magnet ramping speed becomes a limiting factor: separation and recombination of the different energy beams could be performed without ramping the magnets, enabling continuous operation. This was demonstrated by CBETA \cite{bartnik_cbeta_2020} using a dedicated splitter section. A more ambitious proposal might be for a future collider, where regions with zero dispersion are vital to maximise luminosity: this is achieved using dedicated splitter and combiner magnets for eRHIC \cite{aschenauer_erhic_2014}. A closed-dispersion arc would enable beam transport of a continuous energy spectrum, unlike the cases of CBETA and eRHIC which are limited to discrete energies.

At the University of Melbourne, the Technology for Ultra-Rapid Beam Operation (TURBO) project seeks to demonstrate technologies for removing the energy layer switching time bottleneck associated with the beam delivery system. This will be achieved using a scaled-down beamline at low energies with an equivalent momentum range to a clinical beam delivery system. By using Fixed Field Accelerator (FFA) optics, where the magnetic field varies spatially rather than temporally, it is possible to deliver the full range of energies much more rapidly than is currently possible. Though there have been many proposals for medical FFAs, this would be the first demonstration of an accelerator for medical applications using fixed fields including zero dispersion regions. 

\section{Fixed Field Accelerators with Zero-Dispersion Regions}

We begin with an overview of the core concepts of FFAs, including the design rationale, tracking codes, and examples that have been constructed. This leads into a summary of FFA proposals with zero-dispersion regions. These ideas are then synthesised and developed into our new design methodology, as applied to the TURBO project.

\subsection{Review of FFAs}

In a particle accelerator, the transverse bending strength is proportional to the local magnetic field and the focusing strength is proportional to its gradient. Both are inversely proportional to the beam momentum, so they are kept constant in most accelerators by ramping up magnetic fields synchronously with the rising beam momentum. The result is that all energies follow the same closed-orbit around the accelerator, but the momentum acceptance is small, usually less than \SI{\pm 1}{\percent}, and the energy switching time is limited by magnet hysteresis. An alternative would be to vary the magnetic field in space, with each energy having a unique trajectory: this is the core concept behind FFAs. This scheme allows for rapid acceleration and a large momentum acceptance, at the expense of larger and more complex magnets.

Fixed Field Accelerators with sector magnets were first proposed in 1956, around the advent of the strong-focusing synchrotron \cite{symon_fixed-field_1956, sessler_innovation_2010}. FFAs have some advantages over synchrotrons or cyclotrons, such as their large energy acceptance and rapid acceleration capabilities. However, difficulties with the beam dynamics and complicated magnet designs have meant that cyclotrons, synchrotrons, and linacs have dominated since the \num{1960}s. More recently, FFAs have experienced a resurgence with the advent of computer-aided design techniques, leading to FFA proposals for muon colliders \cite{ahdida_nustorm_2019}, LHC successors \cite{trbojevic_permanent_2021}, and often, as the main accelerators for medical facilities \cite{peach_conceptual_2013, garland_normal-conducting_2015, meot_raccam_2019}. They have also been proposed for beam transport and delivery for medical applications, the subject of this work.

In general, the beam dynamics in an FFA is a function of momentum, as the focusing strength varies with energy. In addition, the particle trajectory is a function of momentum, meaning each energy has a unique closed orbit and experiences a different magnetic field. It is possible to enforce a phase advance that is constant by definition by having a magnetic field that follows a `scaling law' \cite{craddock_cyclotrons_2008}. Today, there are many variations of the `scaling law', depending on whether the beam moves in the horizontal or vertical plane as momentum varies, and whether the beamline is straight or follows a circular path: the best known is for the horizontal arc, where the magnetic field in the horizontal midplane is 

\begin{equation}
    B = B_0\left(\frac{r}{r_0}\right)^k \mathcal F \left(\theta\right),
    \label{eqn: FFA_scaling}
\end{equation}

where $r_0$ and $B_0$ are a reference radius and field respectively, the field index $k$ gives the power law dependence of the field, and $\mathcal F$ describes how the magnetic fringe and body fields vary around the machine circumference. The scaling law ensures constant tunes but requires that all closed orbit trajectories are scale enlargements of one another, restricting the accelerator optics. This leads to a large difference between the trajectories of low and high energy beams (known as orbit excursions). As a consequence, the dispersion in a scaling FFA is a nonzero constant. Strictly speaking, the scaling law is only valid where the accelerator has sector-shaped magnets following the curvature defined in Equation \ref{eqn: FFA_scaling}: if rectangular magnets are used, the accelerator can only be approximately scaling. However, the difference between exactly and approximately scaling FFAs has been shown to be negligible where the curvature is sufficiently low \cite{peach_conceptual_2013}. FFAs which do not adhere to the field in Equation \ref{eqn: FFA_scaling} are generally referred to as `nonscaling'.

Most accelerator modelling codes are not suitable for modelling FFAs, as they assume small angles and energy variations to simplify beam tracking. There are several specialised codes that are able to reproduce accurate dynamics for FFAs, each with different capabilities and weaknesses. In this work, Zgoubi \cite{meot_ray-tracing_1999} is used for particle tracking, as it enables sophisticated control over the lattice parameters and collective effects can be neglected in this case. The Python interface Zgoubidoo \cite{vanwelde_zgoubidoo_2023} is used to simplify the simulation inputs, enabling the use of optimisation and analysis tools more sophisticated than those included in a pure Zgoubi environment.

Several FFAs have been successfully constructed and operated. In Japan, a scaling FFA complex has been in operation for many years \cite{tanigaki_present_2006, suga_remodeling_2019}, used for basic accelerator science experiments as well as investigations into Accelerator Driven Subcritical Reactors, and more recently including energy recovery with an internal target for radioisotope production \cite{okita_beam_2019}. Outside of Japan, there have been two nonscaling FFAs: EMMA \cite{machida_acceleration_2012} demonstrated that rapid resonance crossing in nonscaling FFAs was not detrimental to beam quality, and CBETA \cite{bartnik_cbeta_2020} showed that the fast acceleration and large momentum acceptance achievable by FFAs works well for Energy Recovery Linacs.

At present, there are four future FFA projects other than TURBO, at various stages of development. In the UK, FFA options are being investigated for a future neutron spallation source as part of the ISIS-II programme, and a high-intensity proton driver FFA will be demonstrated with the Front End Test Stand (FETS) at RAL \cite{machida_ffa_2023}. At Jefferson Lab, the CEBAF upgrade project may use an FFA to increase the maximum energy from \SI{12}{\giga\electronvolt} to \SI{22}{\giga\electronvolt} by reusing the same tunnel space but increasing the energy acceptance in the recirculating arcs \cite{benesch_cebaf_2023}. For medical applications, a heavy ion research facility is being investigated in America \cite{johnstone_new_2023}, proposing to use an isochronous `racetrack' FFA, and an FFA with spiral sectors has been proposed for the second stage of the LhARA accelerator complex \cite{aymar_lhara_2020} to accelerate protons up to \SI{127}{\mega\electronvolt} for radiobiological research. The principal focus of these projects is on beam acceleration, whereas the TURBO project is investigating beam transport and delivery in more detail.

\subsection{Closed-Dispersion Arc Considerations}\label{subsec: considerations}

One of the major challenges with using an FFA arc for particle therapy is that the particle trajectory is a function of energy, unlike in synchrotrons where the ramping fields ensure that the ideal particle path does not change. To enable rapid energy variation, all particle trajectories must converge (in both position and angle) at either end of the beamline. In the linear matrix formalism, the position of a particle in the horizontal plane with momentum deviation $\delta$ is given by \cite{wiedemann_particle_2015}

\begin{equation}
    \begin{pmatrix}
    x\\
    x'\\
    \delta
    \end{pmatrix}_2
=
    \begin{pmatrix}
    R_{11} & R_{12} & D \\
    R_{21} & R_{22} & D'\\
    0      & 0      & 1 
    \end{pmatrix}_{(2|1)}
    \begin{pmatrix}
    x\\
    x'\\
    \delta
    \end{pmatrix}_1 ,
\end{equation}

where $(D, D')$ is the dispersive element. To comprise a `closed-dispersion arc', both the dispersion and its derivative must be zero at both ends of the beamline.

There are many dispersion suppressor schemes, including half or missing-bend optics \cite{holzer_lattice_2014} where the transition to a zero-dispersion section can be completed smoothly by insertion of matching cells: although these methods have been developed for synchrotrons, there is no reason why the first order results should not also apply to FFAs. In the half-bend scheme, the requirements are that: the bend angle of the suppressor is half that of the normal optics; the lattice is periodic; $D'$ is zero at the interface between the normal and suppressor optics; and the dispersion suppression region must have a phase advance of $(2n+1)\pi$, where $n$ is an integer. 

Large energy acceptance beam transport systems with fixed fields have been explored previously. If the beam to be transported is composed of several discrete energies, it is possible to use splitter magnets which use dipoles and quadrupoles to separate discrete energies and manipulate path lengths and lattice parameters to allow for specific matching for each one. This was successfully demonstrated in CBETA \cite{bartnik_cbeta_2020} and has been proposed for CEBAF \cite{bodenstein_designing_2023}, however it would not be suitable for particle therapy as the energy range is continuous. Another technique employed by CBETA to bring the beam dispersion to zero is to create an `adiabatic transition', where the dipole gradient around an arc is gradually reduced following a sigmoid curve, such that the final lattice has no bending and the overall beam has almost zero dispersion. This has been proposed for particle therapy with a potential design \cite{dascalu_beam_2021}, however the need for the bending strength variation to be slow along the beam path length requires a long beamline with many unique magnets, which would be more expensive and potentially unreliable in a clinical environment. 

Other methods of creating a closed-dispersion arc have been devised, that do not use adiabatic transitions or discrete splitter arcs. Where beamline footprint reduction is paramount, the large bending angle per magnet has been found to require the addition of sextupolar and higher order multipoles to ensure strong focusing over the full range of momenta \cite{zhao_design_2020, nesteruk_large_2019}. In these cases, to keep the beam excursion range small, the momentum acceptance is generally limited to less than $\pm$\SI{15}{\percent}. Conversely, in designs where the bending strength is reduced, combined-function dipole/quadrupole magnets with fine-tuned length, strength, and edge angles can transport all energies \cite{pasternak_novel_2013, brouwer_achromatic_2019, trbojevic_superb_2021}. However, such delicate fine-tuning may be sensitive to errors, and the difficulty of building a lattice with many unique magnets and a high packing factor has not been explored. The use of scaling FFA optics has also been considered for beam delivery at a medical facility \cite{tesse_gantry_2023, tesse_achromatic_2023}, where the upsides of the straightforward beam optics are counteracted by limitations imposed by the scaling law, such as large beam excursions. By combining nonscaling multipolar fields with low-dispersion beamlines, the energy acceptance can be increased while limiting the transverse beam excursion range \cite{machida_beam_2010}.

With a scaling FFA, the constant tune for all energies appears to make the design of a dispersion suppressor with a large energy acceptance no different to the synchrotron case. However, particle trajectories in a scaling FFA are constrained by the natural dispersion which is required to keep the focusing strength constant over the full range of energies: even though the optics imply zero dispersion, the closed-orbit trajectories do not. Previous work  has indicated that scaling FFA optics can provide a good starting point for a dispersion suppressor \cite{machida_beam_2010}, however it is limited by the chromatic aberrations arising from particles not following the periodic trajectories. By breaking the scaling law, the effectiveness of the dispersion suppressor may be improved \cite{fenning_high-order_2012}, although it can lead to severe distortions in the final beam where the CS parameters are not correctly matched.

We define momentum acceptance $\alpha_p$ as

\begin{equation}
    \alpha_p = \frac{p_\text{max}-p_\text{min}}{p_\text{max}+p_\text{min}},
\end{equation}

such that the reference momentum $p_0$ is given by the middle of the momentum range, and related to the maximum and minimum momenta by

\begin{align*}
    p_\text{min} &= p_0(1-\alpha_p), \\
    p_\text{max} &= p_0(1+\alpha_p) . \\    
\end{align*}

As only protons are considered in this study, the rigidity and momentum acceptances are equivalent. For the TURBO project, the momentum acceptance is \SI{\pm 42}{\percent}.

\section{Halbach Arrays without Custom Magnets}\label{sec: magnets}

Fixed Field Accelerators utilise magnets where the fields do not need to vary in time. These magnets are usually combined function, as FFAs require a high packing factor to achieve sufficient focusing. FFAs thus break both the assumption of separated-function magnets and of ramping fields, in contrast to most accelerators where the separated function magnetic fields of dipoles, quadrupoles and higher order correctors must be ramped simultaneously as the particle energy increases. In all cases, if the required fields are below \SI{1.8}{\tesla} this can be achieved with electromagnets, otherwise superconducting magnets may be required. An advantage of using static magnetic fields is to enable a third option: so long as the required fields are sufficiently low, permanent magnets can be used to create the required multipoles. Here we further consider the use of commercially available permanent magnets to reduce overall costs and waste, while allowing rapid prototyping and design.

Permanent magnet arrays have been employed in accelerators for several decades \cite{shepherd_permanent_2020}. With the advent of rare earth magnets in the 1980s, Halbach proposed that correctly oriented blocks of permanent magnet material could be used to produce solenoidal, helical, and multipolar fields \cite{halbach_design_1980}. They are often used in light sources, wigglers and undulators made with permanent magnets can be moved together or apart to tune the synchrotron radiation \cite{clarke_science_2004}. Halbach quadrupoles are also used for linacs, where the energy at a given point does not change \cite{thonet_use_2016, wangler_rf_2008}. A recent innovation for conventional synchrotrons is a tunable permanent-magnet: the ZEPTO project allows for varying the field strength of permanent magnets by moving the magnetic material relative to an iron yolk \cite{shepherd_tunable_2014}. For Fixed Field Accelerators, magnet arrays can be constructed without requiring momentum-dependent tunability, allowing for more specialised permanent magnet designs.

By arranging blocks of permanent magnet material, it is possible to produce arbitrary combined function multipoles. This was trialled with CBETA \cite{brooks_cbeta_2019}, which required combined function dipole/quadrupole magnets in some sections of the return arc. For each of these magnets, \num{16} custom trapezoidal wedges of NdFeB were used to approximate the desired field: each wedge required a unique size and magnetisation direction, which dominated the cost of the magnet arrays. It was found that multipole errors arising from assembly could be corrected with thin iron rods, inserted into the magnet bore following a correction algorithm. A similar permanent magnet solution is being explored for the CEBAF upgrade \cite{brooks_permanent_2022}.

A different method of designing permanent magnet arrays is being investigated for the TURBO project. Rather than using expensive custom permanent magnet pieces, commercially available magnet blocks can be combined to produce the desired fields to sufficient accuracy. As well as lowering costs, using many identical magnetic blocks ensures that all the magnetic material can be readily reused for other projects, which would not be possible if each magnet required custom-designed pieces. Although this method requires a larger total volume of magnetic material -- due to the gaps in the magnet introduced by the segmentation -- we estimate that the total cost is an order of magnitude less than for custom blocks. One concern is that the field quality of the magnet array may be worse than with custom magnets: in Section \ref{sec: robustness}, we show that the field quality is sufficient for the demonstrator beamline where only a single pass through each magnet is necessary.

We demonstrate the magnet design algorithm with a modified version of the magnet denoted `BDT1' in \cite{brooks_permanent_2020}, which was developed for the CBETA return arc. This is a combined function dipole/quadrupole with strengths of \SI{{-}0.1002}{\tesla} and \SI{11.1475}{\tesla/\meter}. The aperture radius of the array used here is approximately \SI{45}{\milli\meter}, with a good field region with a \SI{15}{\milli\meter} radius. The magnets used for CBETA were \SI{122}{\milli\meter} long, however we have shortened our example to \SI{100}{\milli\meter} to match the requirements of the TURBO project. The individual magnet blocks are modelled as $\num{12.7}\times\num{12.7}\times\SI{100}{\milli\meter^3}$ cuboids with a remanent  field $B_r$ of \SI{1.3}{\tesla}, as this matches commercially available products such as those available from \cite{noauthor_100mm_2024}. Magnet modelling was performed using the Python package Magpylib \cite{ortner_magpylib_2020}, which allows for fast optimisation but does not include demagnetisation in the presence of strong B-fields. 

\subsection{Magnet Array Design Methodology}\label{subsec: mag_design}

To generate a magnetic multipole of order $N$ with idealised bulk permanent magnet material, the continuously-varying magnetisation in plane-polar coordinates $(\rho, \phi)$ follows

\begin{equation}\label{eqn: halbach_magnetisation}
    \vec M = M_0 \left[
    \cos\left(N\left(\phi-\phi_0\right)\right)\hat\rho +
    \sin\left(N\left(\phi-\phi_0\right)\right)\hat\phi
    \right],
\end{equation}

where reference angle $\phi$ is $\pi/2$ or $3\pi/2$ for normal multipoles, or a multiple of $\pi$ for skew multipoles \cite{halbach_design_1980}. The field strength inside the magnet bore is determined by the inner and outer radii of the cylinder ($r_i$ and $r_o$ respectively) and the magnet remanence field $B_r$. For infinitely long N-poles, the field inside the bore is given by

\begin{align}
    \vec B &= B_r \ln\left(\frac{r_o}{r_i}\right)\hat y  &&N=1,
    \label{eqn: halbach_field_0}\\
    \vec B &= B_r\left(\frac{\vec x + \vec y }{r_i}\right)^N \frac{N}{N-1} \left[1-\left(\frac{r_i}{r_o}\right)^N\right] &&N \geq 2, 
    \label{eqn: halbach_field_1}
\end{align}

which suggests that for fixed cylinder dimensions, multipole strength decreases with increasing order.

\begin{figure*}[]
    \centering  
    \includegraphics*[width=0.8\textwidth]{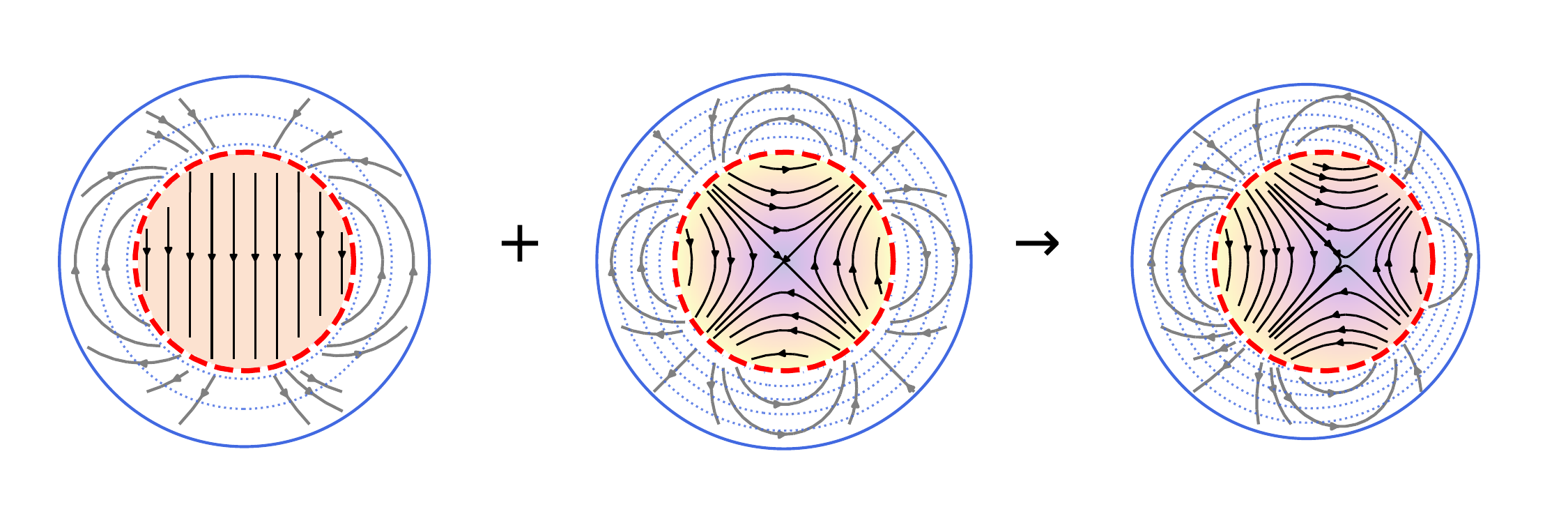}
    \caption{Combining dipole and quadrupole fields to produce a combined-function magnet. The B-field is shown within the dashed red circles, and the ideal magnetisation vectors are given by the exterior grey streamlines. The equipotentials of the magnetisation vectors are shown in blue: for pure N-poles they are circular, but for the combined-function case they are distorted. The distorted streamlines and contours in the combined-function magnet are from addition of the magnetisation vectors.}
    \label{fig: halbach_superposition}
\end{figure*}

Using Equations \ref{eqn: halbach_magnetisation}--\ref{eqn: halbach_field_1}, the magnetisation required to produce any individual N-pole can be calculated. However, it is not immediately apparent how to combine the magnetisation vectors to create an arbitrary combined-function multipole.  From the design of superconducting magnets \cite{bottura_superconducting_2016}, it is known that the field for a multipole of order $N$ is produced inside a longitudinal $\cos(N\phi)$ current distribution. In addition, the field outside this current ring corresponds to the magnetisation that would produce the N-pole, due to the magnetic field reciprocation theorem \cite{jeans_mathematical_1911, prat-camps_circumventing_2018}. As such, by superposing a series of $\cos(N\phi)$ current rings, we can simultaneously produce any combined-function magnetic field and the ideal continuously varying magnetisation that would produce it. From this we can derive a layout of permanent magnet segments to produce the correct magnetisation. An illustrative example is shown in Fig.~\ref{fig: halbach_superposition}. 

Our design procedure is as follows: we first use the cosine-theta current rings to produce the internal fields, normalised to the desired strengths; we then apply linear superposition of magnetic fields to combine the fields outside the rings, which is the dual of the desired magnetisation. In Fig.~\ref{fig: halbach_superposition}, we see that the magnetisation follows Equation \ref{eqn: halbach_magnetisation} for the N-poles, and is a linear combination for the combined-function magnet. The equipotential lines indicate the outer distance of magnetic material required to produce a magnetic field of a desired strength. An advantage of this method over just using Equation \ref{eqn: halbach_magnetisation} is that it calculates both the field and the magnetisation simultaneously, allowing both to be used in optimisation. The next challenge is to find an arrangement of permanent magnets which approximates the ideal solution found here. 

The goal of the optimisation algorithm is to reproduce a set of multipoles using an array of discrete magnet blocks. This is quantified by the field quality of a magnet, giving the difference between the design and achieved multipoles. Field quality is defined by the Fourier expansion of the radial magnetic field at a radius $r$ \cite{russenschuck_design_2022},

\begin{align}
    B_n &= \frac{1}{\pi}\int_{-\pi}^\pi B_r(r, \phi)\sin(N\phi) d\phi \label{eqn: B_n}, \\ 
    A_n &= \frac{1}{\pi}\int_{-\pi}^\pi B_r(r, \phi)\cos(N\phi) d\phi \label{eqn: A_n},
\end{align}

where the $B_N$ and $A_N$ correspond to the normal and multipole components respectively. In many cases, these are normalised relative to a `main' multipole component (usually quadrupole). The stray multipole terms in an accelerator are usually required to be less than one part in \num{E4} of the main component.

For both optimiser stages described here, we define $B_{N, \text{opt}}$ as the goal multipoles for some magnet, and $B_{N, i}$ as the multipoles from a possible solution. The objective function $f$ to be minimised is given by

\begin{equation}\label{eqn: mag_obj_func}
    f=\sqrt{ \sum_N  \left(B_N-B_{N, \text{opt}}\right)^2+\left(A_N-A_{N, \text{opt}}\right)^2},
\end{equation}

which is constrained by keeping all magnets outside the beampipe, and to prevent intersections. Multipoles are calculated at the longitudinal midpoint of the magnet at the outer radius of the good field region $(r=R_\text{good})$.

In cases where the discrete magnet blocks are not able to fully reproduce the desired set of multipoles, it may be preferred to use the midplane magnetic field as the optimisation objective. This is particularly useful where the horizontal beam excursion is much larger than the vertical size of the beam, as is often the case in FFAs. The multipole error inside the good field region may increase if the optimisation objective is the midplane field, however the greatest error is beyond the vertical beam extent. The midplane field error to be optimised is

\begin{equation}\label{eqn: midplane_error}
     \left.
     \sum_{i=0}^{p}   
     \bigg(B_y(x_i)-B_{y, \text{opt}}(x_i)\bigg)^2 
     \right \vert_{\shortstack{y=0\\z=0}},
\end{equation}

where ${x_i}$ spans the good field region, with $x_0= -r_\text{good}$, $x_p= r_\text{good}$. The horizontal B-field can be neglected as the enforced symmetry of the magnet arrangements keeps skew multipoles small, allowing us to exclude $B_x$ in this calculation.

\begin{figure}[]
    \centering  
    \includegraphics*[width=0.85\columnwidth]{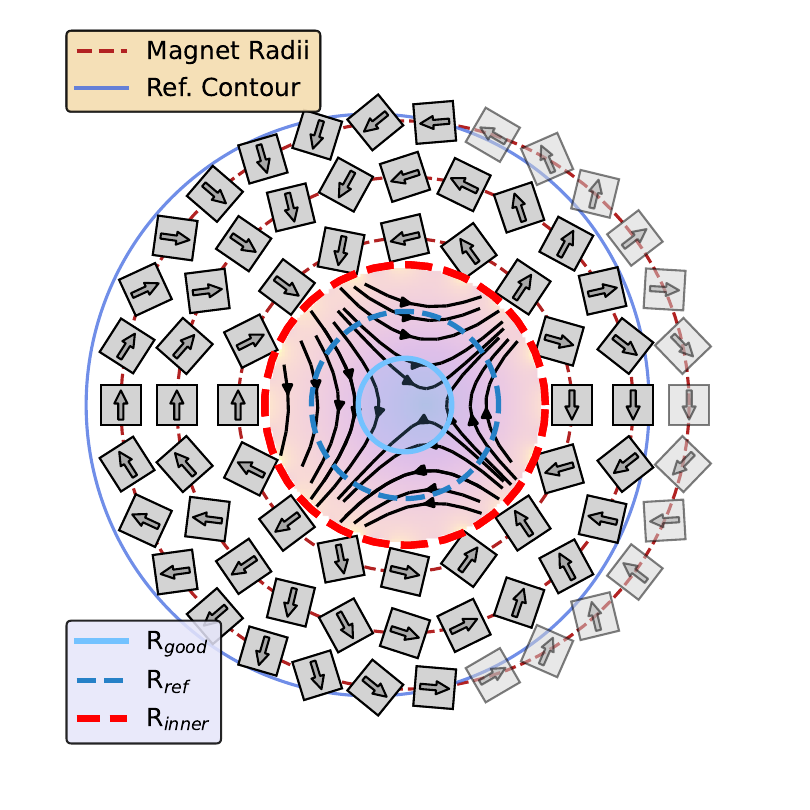}
    \caption{The first stage in the magnet optimisation algorithm, where the variables are the radii for magnet placement and the outermost contour from Fig.~\ref{fig: halbach_superposition}. Magnets are transparent if they are outside the reference contour: they are excluded from the field calculation. The good field region and reference radius for multipole calculations are shown.}
    \label{fig: mag_opt_stg_1}
\end{figure}

The optimal arrangement of magnetic blocks, to first order, will follow the contour lines shown in Fig.~\ref{fig: halbach_superposition}, oriented such that their magnetisation follows the local field direction. Our algorithm approximates this by placing magnet blocks in concentric layers, and removing any that lie outside the ultimate contour defined by some `reference potential'. An optimisation algorithm -- in this case, simulated annealing -- then varies the radii of the layers and the value of the reference potential, minimising the objective function. The number of magnets per layer is maximised to ensure the strongest possible B-field. The first step in the optimisation process is shown in Fig.~\ref{fig: mag_opt_stg_1}.

Although this initial optimisation is usually good at finding an approximate solution for the magnet placement, it is not always successful. One issue can be attempting to use too few or too many magnet layers: if there aren't enough magnet layers, it will not be possible to make a strong enough magnetic field inside the magnet bore; if there are too many rings, the algorithm execution time can become prohibitive, and the extra search space makes it difficult to converge on a good result. Assuming the dipole term is dominant, the approximate number of magnet rings can be calculated using Equation \ref{eqn: halbach_field_0}, and can be fine-tuned if the algorithm fails to converge.

\begin{figure}[]
    \centering  
    \includegraphics*[width=0.85\columnwidth]{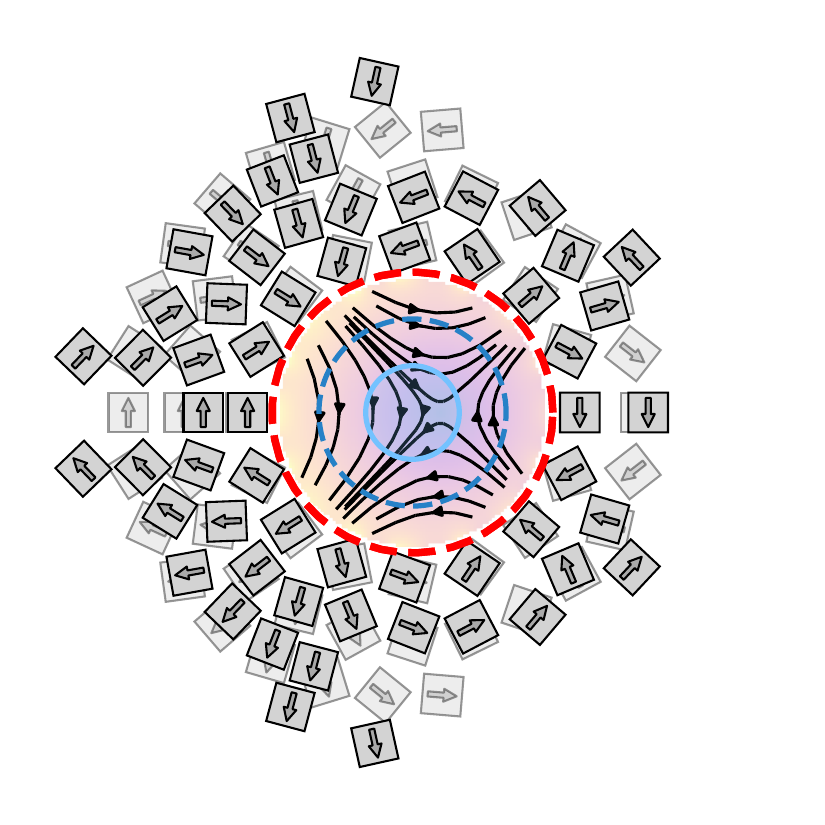}
    \caption{The second magnet optimisation step. The result from the first stage, used as the initial points, is shown in the background. The good field region (shaded blue region) and reference radius (dashed blue circle) are indicated for comparison with Fig.~\ref{fig: mag_opt_stg_1}.}
    \label{fig: mag_opt_stg_2}
\end{figure}

The second optimisation step searches for a minimum in the vicinity of the solution from the first stage. For every magnet, the horizontal and vertical positions can be varied, as well as the rotation angle. This significantly increases the number of optimisation variables, and as such, the second optimisation stage takes longer to execute than the first. An example array from the second stage of optimisation is shown in Fig.~\ref{fig: mag_opt_stg_2}. The number of variables is reduced by noting the midplane symmetry of the B-fields required for accelerator applications: only the magnets in the `top-half' of the magnet array need to be optimised, as the bottom half can be found by mirroring magnets through the array midplane and reflecting each magnet about the vertical axis. In addition, the magnet positions and orientations are constrained to prevent non-physical overlaps, with some extra space given to ensure that the magnet mount will be feasible from an engineering perspective.

\begin{figure}[]
    \centering  
    \includegraphics*[width=0.95\columnwidth]{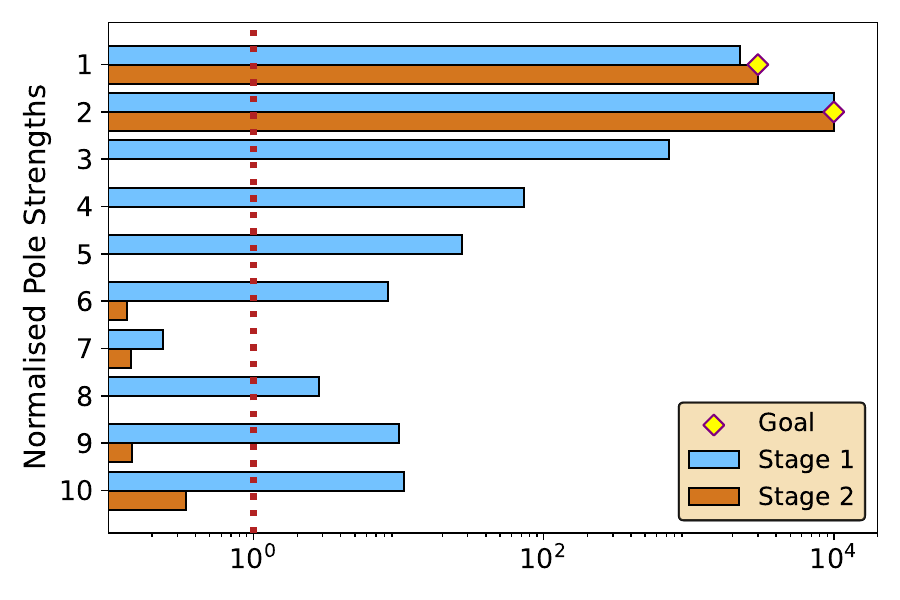}
    \caption{Pole strengths after both optimisation stages, relative to the ideal quadrupole value.  
    The second stage brings all undesired multipoles below one part in $10^4$. Only normal poles are shown, as skew poles are negligible due to symmetry.}
    \label{fig: mag_opt_multipoles}
\end{figure}

After the second algorithm stage, the solution can be quite different to the first arrangement as shown in Fig.~\ref{fig: mag_opt_stg_1}, due to the gaps introduced between the magnets. Despite this, we see in Fig.~\ref{fig: mag_opt_multipoles} that the magnitude of the error multipoles has gone down significantly during the second optimisation stage. It is not possible to say whether the solution displayed in Fig.~\ref{fig: mag_opt_stg_2} is globally optimal given our constraints, as the parameter space of possible magnet arrangements is too large to fully explore.

\begin{figure*}[]
    \centering  
    \includegraphics*[width=0.9\textwidth]{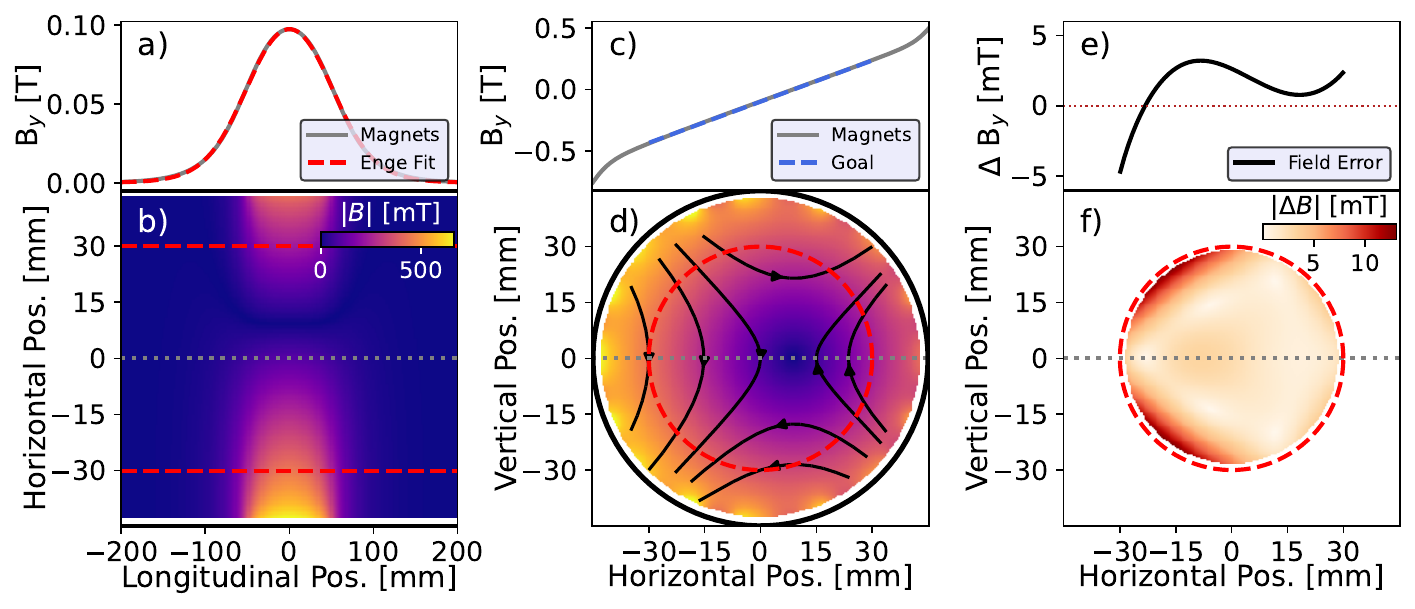}
    \caption{Magnetic fields achieved with our optimised magnet array. The black and red circles in the lower plots show the magnet array inner radius and the internal reference radius respectively.  \textbf{(a, b)} Magnetic field strength in the horizontal midplane along the longitudinal axis. \textbf{(c, d)} Magnetic field in the transverse plane measured at the centre of the magnet. \textbf{(e, f)} Error in achieved magnetic field at the magnet centre, relative to the ideal field.}
    \label{fig: all_achieved_fields}
\end{figure*}

The magnetic field produced by this magnet array is explored further in Fig.~\ref{fig: all_achieved_fields}. Even though the longitudinal magnet edges only extend to \SI{\pm 50}{\milli\meter}, the fringe fields extend far beyond this limit, with a small residual field up to \SI{100}{\milli\meter} beyond the magnet ends. These long fringes are a consequence of the large inner bore radius relative to the magnet length: in a closed-dispersion arc with long magnets or a small magnet bore, the fringe fields could be significantly reduced. In the transverse plane at the magnet centre, the desired multipolar fields are accurately reproduced, although the field error grows at larger radii. Interestingly, we see in Fig.~\ref{fig: all_achieved_fields}e) and ~\ref{fig: all_achieved_fields}f) that the smallest error is at a radius of approximately \SI{15}{\milli\meter}: this corresponds to the radius of the `good field region' set earlier, suggesting that the judicious choice of the good field region is important to ensure the optimisation routine is effective. 

\subsection{Modelling Magnets for Tracking Studies}\label{subsec: mag_modelling}

During the beam optics design phase, we must construct a representation of the magnetic field that is suitable for rapid calculations and optimisation. Though it is possible to use a full 3D fieldmap in Zgoubi, it is not practical for all design simulations as each iteration would require creation and optimisation of a new magnet array. In addition, particle tracking through a full fieldmap in Zgoubi is generally slower than using in-built components, due to the extra computation required for interpolation.

Magnetic elements in Zgoubi can be modelled using multipole coefficients, specified by the field at the magnet inner bore of each contributing multipole. To account for the magnetic field beyond the magnet edges, Zgoubi uses the Enge-type fringe fields \cite{enge_deflecting_1967}. Zgoubi also has options available to modify the fringe fields as a function of the horizontal offset from the magnet centre, however this is not required for this model as the magnet bore is cylindrical. The Enge fit works well in our case, as seen in Fig.~\ref{fig: all_achieved_fields}a), confirming that the Enge model is suitable for Halbach arrays. The Enge coefficients are found with a least-squares fit, measured along the longitudinal axis of the magnet. Although the fitting parameters will be different for each magnet, it is reasonable to assume that they are all approximately the same for the initial modelling of the TURBO beamline and varying them in subsequent iterations.

As the magnet packing factor for the TURBO beamline will be large, the fringe fields will overlap between magnets. This is not possible using most Zgoubi elements, however overlapping fringes are an option when the \texttt{DIPOLES} element is used. Although this is intended for modelling sector magnets, it was demonstrated for the PAMELA project \cite{sheehy_fixed_2010} that the sectors could be `straightened' by scaling up the effective radius, and reducing the effective angular extent by the same amount. This enables the use of straight magnets with overlapping fringe fields in Zgoubi, so long as no more than \num{5} magnets are required per section. For the TURBO project, we employ this method to model cells with three magnets, assuming that the fringe field overlap between cells is negligible. 

\subsection{Field Sensitivity to Magnet Block Errors}

It is important to consider how errors in the individual magnet blocks contribute to field imperfections in the overall array. These errors can be from misplacements of the individual magnets in their mount, or from misalignments of the magnetisation axis from the expected direction. Very large errors, such as a full \SI{90}{\degree} rotation of an individual magnet block, should be discovered during fabrication and as such only small errors that are difficult to correct are discussed here.

We define a Figure of Merit (FoM) based on Equation \ref{eqn: mag_obj_func}, modified to

\begin{equation}\label{eqn: mag_fom}
    \sqrt{ \sum_{N=1}^{15}  \left(b_N-b_{N, \text{mag}}\right)^2+\left(a_N-a_{N, \text{mag}}\right)^2},
\end{equation}

where $(b_N, a_N)$ represent the normalised $N$-order normal and skew multipoles from a given configuration, and $(b_{N, \text{mag}}, a_{N, \text{mag}})$ are found from our stage 2 output from Section \ref{subsec: mag_design}. In this case, all the values are normalised to the quadrupole component of the ideal magnet. Multipole values are normalised in the same way as in Fig.~\ref{fig: mag_opt_multipoles}, such that the normal quadrupole coefficient $b_2=$ \num{E4}. For CBETA, the average FoM was \num{41.09} before tuning \cite{brooks_permanent_2020}, although values over \num{100} were recorded in some cases. Our FoM is expected to be slightly larger than the CBETA values, as their calculation does not include the dipole contribution, but the two are otherwise directly comparable.

The most probable error in magnetisation is due to misalignment of the physical axes of the magnets and the magnetisation direction. So far, it has been assumed that the magnetisation is exactly aligned with the magnet, however errors may be introduced. Errors in the magnetisation are more difficult to detect than magnet misalignments, which should be discovered by visual inspection. To study the impact of magnetisation errors, we assume that the the error is entirely in the transverse plane (i.e. no longitudinal magnetisation), and that the overall remanant field $B_r$ is not changed. The error angle standard deviation is varied up to \SI{1}{\degree}. Each individual magnet is assigned a normally distributed error, and \num{1000} simulations are performed for each error magnitude.

\begin{figure}[]
    \centering  
    \includegraphics*[width=0.95\columnwidth]{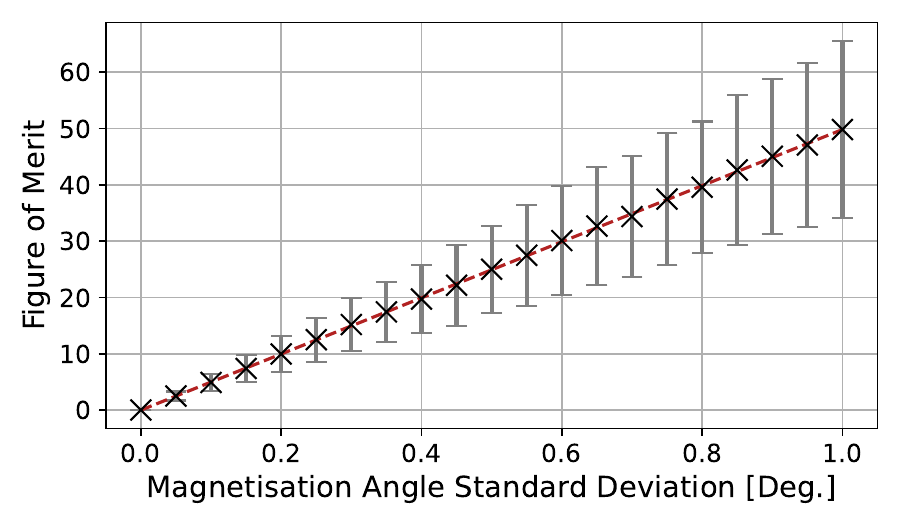}
    \caption{Error in magnetisation angle. The Figure of Merit is defined by Equation \ref{eqn: mag_fom}. Studying magnetisation errors up to \SI{1}{\percent}, the FoM varies linearly with error magnitude.}
    \label{fig: mag_dir_error}
\end{figure}

In Fig.~\ref{fig: mag_dir_error}, we see that the FoM grows linearly with the error magnitude, with a gradient of \SI[per-mode=symbol]{49.8 \pm 0.2}{\per\degree}. If our magnet tuning regime has the same requirements and success as CBETA's, this suggests that magnetisation errors up to \SI{2}{\degree} would be acceptable, in the absence of other errors. 

The other error that may arise will be due to misalignments of the magnet blocks, in both position and angle. These errors will depend on the precision that can be reached with our magnet mounts: here we set the maximum position error in the horizontal plane to \SI{100}{\micro\meter}, and rotation about the longitudinal axis \SI{1}{\degree}. These values may be difficult to reach if the magnet mounts are 3D printed, but should be feasible if a standard machining approach is used i.e. milled aluminum. In a separate study, we studied the impacts of both position and orientation errors to the FoM: we found that the contribution from the position errors (up to \SI{0.1}{\milli\meter}) is smaller than from the rotational errors (up to \SI{1}{\degree}). The FoM increases linearly with the errors in all cases, suggesting that no single type of error will dominate the final FoM.

In practice, all the different errors considered here will be present, as well as others including temperature-dependent field changes and demagnetisation due to radiation damage. The impacts of irradiation and temperature variation on accelerator magnets has been summarised in \cite{shepherd_radiation_2018}. The magnet arrays will also require good alignment relative to the beamline, which should be achieved using standard metrology techniques. It is expected that the short length of this single-pass beamline should ensure that the errors will not significantly degrade beamline performance: this hypothesis is explored  in Section \ref{sec: robustness}.

\subsection{Limitations of Halbach Arrays without Custom Magnets}

Although the magnet array creation algorithm presented here is sufficient for our first-order design, it has some limitations in its current form. Some of these are due to the specific implementation of the magnet modelling code, and may be solved by using a more complete program such as the finite element analysis code OPERA. One such issue is that the algorithm does not include demagnetisation effects, which may be an issue as there will be strong fields over some of the magnet blocks, although NdFeB is usually robust against demagnetisation. In addition, the required calculation times can become large when several rings are required for the first optimisation stage, but this may be alleviated with a more efficient implementation or more powerful computing resources.

Though the design algorithm performs well, there are opportunities for further improvement. At present, it only allows for one type of magnet for the ring, although it would be possible to use larger magnet blocks to produce the bulk of the field, and smaller blocks to fill in the spaces in between. The requirement for only one type of block makes some magnets infeasible: an example is the CBETA `BDT2' magnet \cite{brooks_permanent_2020}, where the space between the array inner radius and the outermost contour for stage 1 of the algorithm may be smaller than the size of magnet chosen. This may be countered by including more than one type of magnet block, although this is not compatible with the current implementation of the first stage of the optimisation algorithm.

This method of producing permanent magnet arrays works well for the TURBO project, where the ability to reuse magnetic material and reduce costs is desirable. However, a different conclusion may be drawn for accelerators with other requirements, such as a more restrictive error tolerance or strict limitations on the transverse magnet size: in such cases, a design method more like the CBETA trapezoidal wedges may be preferred.

\section{Large Momentum Acceptance Beamline Design}\label{sec: beamline}

Closed-dispersion arcs with a large momentum acceptance for medical applications have been proposed in the past, however one has never been constructed. We present a method for designing such an arc, which could be applied to a particle therapy beamline or other facilities. We use our design methodology on the TURBO technology demonstrator arc, using magnetic fields that could be achieved by the approach in Section \ref{sec: magnets}. Realistic field errors are then considered in Section \ref{sec: robustness}. Although a possible matching section, beam diagnostics, and the beamline end station constitute the final overall design of TURBO, they are beyond the scope of this study.

TURBO must demonstrate transport for \num{0.5}--\SI{3.0}{\mega\electronvolt} protons, a momentum acceptance of \SI{\pm 42}{\percent} (a rigidity range of \num{0.1}--\SI{0.25}{\tesla\meter}). The arc must have a nonzero bending angle to represent realistic beamline requirements, while also fitting within the existing infrastructure in the Pelletron lab at the University of Melbourne. The constraints on the arc design are given in Table \ref{tab:turbo_geometry}.

\begin{table*}[]
    \centering
	\caption{TURBO beamline constraints, from the geometric requirements of the University of Melbourne Pelletron lab, cost of parts, expected feasibility, and safety concerns}
	\label{tab:turbo_geometry}
	\begin{tabular}{rrll}
		\toprule
		\textbf{Parameter} & \textbf{Value} & \textbf{Units}    & \textbf{Explanation}                                               \\
		\midrule
	    Bend Angle         & \num{30}       & \si{\degree}      & Must be nontrivial for this demonstrator arc, constrained by space \\
	    Overall Footprint  & \num{2}        & \si{\meter^2}     & Limited by available space in the Pelletron lab in Melbourne       \\
	    Beampipe Radius    & \num{31.75}    & \si{\milli\meter} & Matches existing Pelletron beampipes                                             \\
	    Magnet Bore Radius & \num{35}       & \si{\milli\meter} & Larger than the beampipe radius, with extra space                  \\
	    Magnet Length      & \num{100}      & \si{\milli\meter} & Matches commercially available permanent magnets                   \\
	    Max B-Field        & \num{0.6}      & \si{\tesla}       & Readily achieved by Halbach arrays, and safer than larger field    \\
     
        \bottomrule
	\end{tabular}
\end{table*}

The TURBO beamline will use beam pipes with DN63LF flanges, which have a \SI{63.5}{\milli\meter} inner diameter and slightly larger outer size: this motivates a maximum beam excursion range of \SI{\pm 25}{\milli\meter} (i.e. \SI{50}{\milli\meter} total), and a magnet inner bore radius of \SI{35}{\milli\meter}. As discussed in Section \ref{sec: magnets}, permanent magnet Halbach arrays will be used to produce the required B-fields. These magnets will be rectangular for ease of fabrication. The maximum magnetic field is set to \SI{0.6}{\tesla}, which should be achievable by the permanent magnets: from Equation \ref{eqn: halbach_field_0}, we find that the outer radius for the magnetic material will be at least \SI{56}{\milli\meter}, although it is expected to be larger due to the additional multipoles and the gaps between magnet blocks. The magnet ends are modelled using Enge-style fringe fields. For the initial arc design, all magnet models in Zgoubi use the same Enge fit coefficients: $C_0=$\num{-3.55E-02}, $C_1=$\num{4.17}, and $\lambda=$\SI{7}{\milli\meter}.

The closed-dispersion arc for the TURBO project will comprise four cells in total. The beam will enter the first cell with both dispersion and its derivative at zero, and must be brought back to the same condition by the end of the beamline. This implies that the dispersion function must be symmetric, with a turning point between the second and third cells. Due to this symmetry, the optics of the third and fourth cells must be mirrored from the second and first respectively, simplifying the arc design. Using the same magnet blocks as in Section \ref{sec: magnets}, we assume that each magnet is \SI{10}{\centi\meter} long. The drift spaces are set as \SI{7.5}{\centi\meter} within the cells  and \SI{11.25}{\centi\meter} between them, preventing extreme overlap of fringe fields and providing space for diagnostics. Each lattice cell is made up of three magnets on a single straight girder: this is modelled in Zgoubi using the methods described in Section \ref{subsec: mag_modelling}. The arc bends a total of \SI{30}{\degree}, with each cell starting and ending with a \SI{3.75}{\degree} rotation of the reference axis. An extra \SI{5}{\centi\meter} drift is added to at the start and end of the arc, representing the additional space that may be included after matching and before the end station: it is likely that this drift length will change in the final version of the TURBO beamline. The layout of the arc, including trajectories for a scaling FFA case, is shown in Fig.~\ref{fig: scaling_disp_supp}.

\subsection{Dispersion Suppression with a Scaling FFA}

As discussed in Section \ref{subsec: considerations} our initial lattice is an FFA following the scaling law. As our case has two cells to suppress dispersion and two to restore it, each one must have a phase advance of $\pi/2$, and the overall arc has a phase advance of $2\pi$. In addition, as we do not need to match to some other periodic lattice, we do not need to halve the bending strength of our suppressor.

\begin{figure}[]
    \centering  
    \includegraphics*[width=0.85\columnwidth]{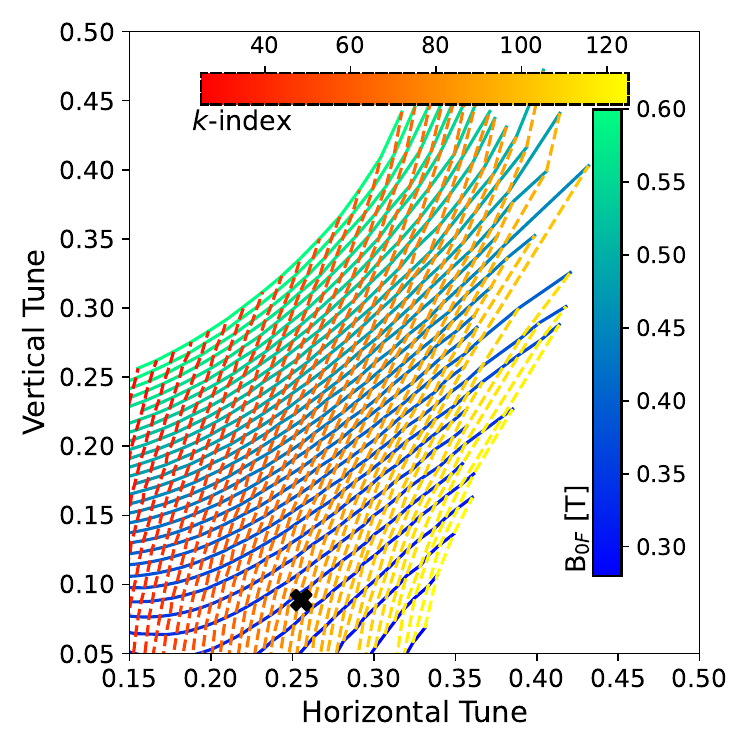}
    \caption{Working point scan for the scaling FFA lattice used as the first iteration of the closed-dispersion arc, by varying the $k$-index and $B_0$ for the focusing magnets. $B_0$ for the defocusing magnets is determined by ensuring the closed orbit of a reference energy starts at the origin. The cross marks the selected working point.}
    \label{fig: choose_working_point}
\end{figure}

A scan of possible working points must be performed to select a scaling FFA suitable for a closed-dispersion arc. For a single cell, the horizontal phase advance must be close to $\pi/2$, which leaves some freedom in the strength of the vertical focusing. The results of the scan of working points for the TURBO geometry is shown in Fig.~\ref{fig: choose_working_point}. The horizontal bending strength $B_{0\text{F}}$ and the $k$-index are varied independently. The value of $B_{0\text{D}}$ is chosen to ensure that the closed orbit passes through the centre of the first magnet in the transverse plane. To first order, the horizontal tune is mostly a function of $k$. The vertical tune rises with increasing $B_{0\text{F}}$. In this case, only the first stability region (i.e. horizontal tunes less than \num{0.5}) is explored, although previous studies have shown that increasing the horizontal focusing into a second stability region can reduce the orbit excursion \cite{machida_scaling_2009} at the cost of tighter magnet tolerances. 

\begin{figure*}
    \centering
    \subfloat[Particle closed orbits through four triplet cells, with the beam rigidity relative to the reference indicated]{\includegraphics[width=0.32\textwidth]{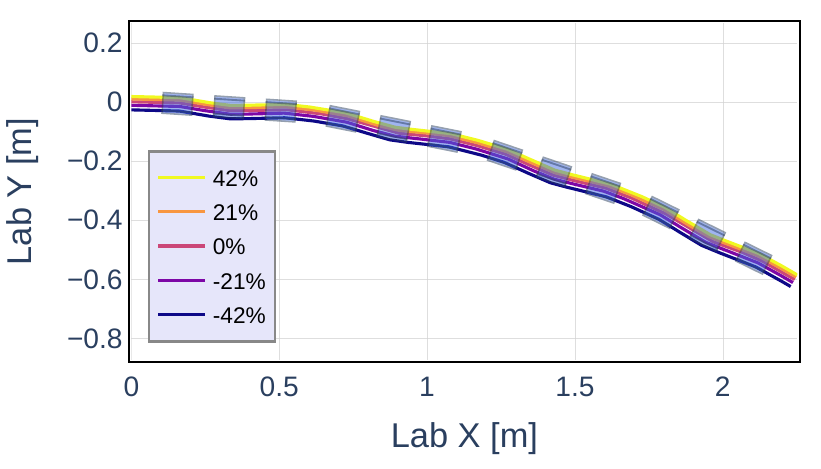}}\label{subfig: scaling_trajectories}%
    \hfill
    \subfloat[Particle trajectories when all momenta are injected at the origin.]{\includegraphics[width=0.32\textwidth]{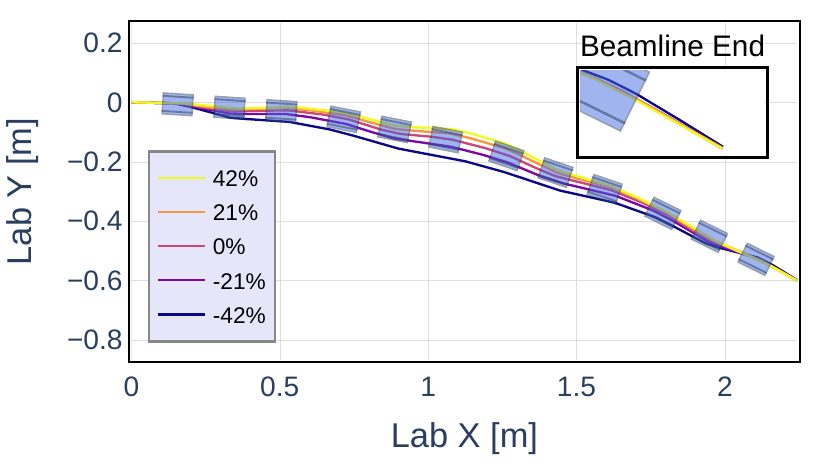}}\label{subfig: scaling_almost_closed}%
    \hfill
    \subfloat[Beta functions for the reference momentum along the closed orbit.]{\includegraphics[width=0.32\textwidth]{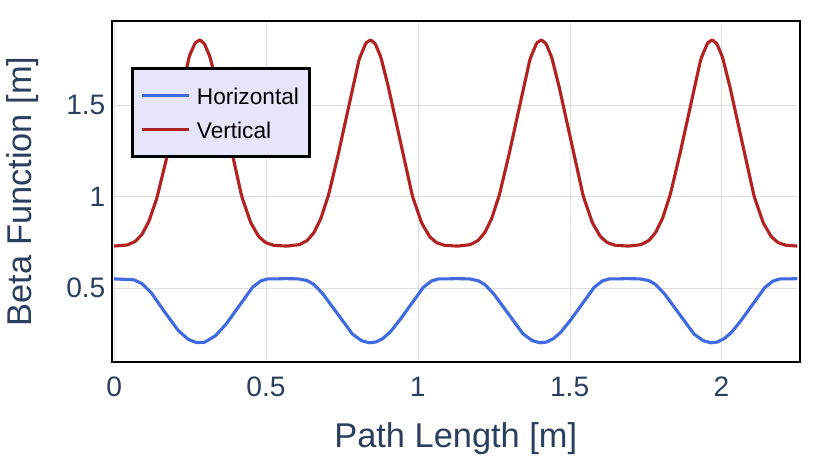}}\label{subfig: scaling_cs}%
        \caption{As the phase advance for the arc is close to  $2\pi$, it is difficult to determine the closed orbit and CS parameters for the full arc with four cells: instead, the calculation is performed for one cell, which is not at such a disruptive resonance.}
        \label{fig: scaling_disp_supp}
\end{figure*}

The phase advance per cell, $k$ value and F/D ratio must be chosen carefully, even prior to full optimisation. To keep the orbit excursion of the scaling FFA small, strong horizontal focusing is required: $k$ must be maximised. However, too large a $k$ value will result in weak vertical focusing, leading to a large vertical beta function which risks increasing the vertical beam size beyond the magnet aperture. In addition, the magnetic fields should not exceed the limit of \SI{0.6}{\tesla}: this limitation excludes much of the high vertical tune space, as well as anything beyond the first stability region. As a compromise between all these requirements, a lattice is chosen with $B_{0\text{F}}= $\SI{0.325}{\tesla}, $B_{0\text{D}}= $\SI{-0.5449}{\tesla}, and $k=$ \num{85}. The closed orbits and Courant-Snyder (CS) parameters of this lattice are shown in Fig.~\ref{fig: scaling_disp_supp}.

\begin{figure}[]
  \centering
  \includegraphics[width=.9\linewidth]{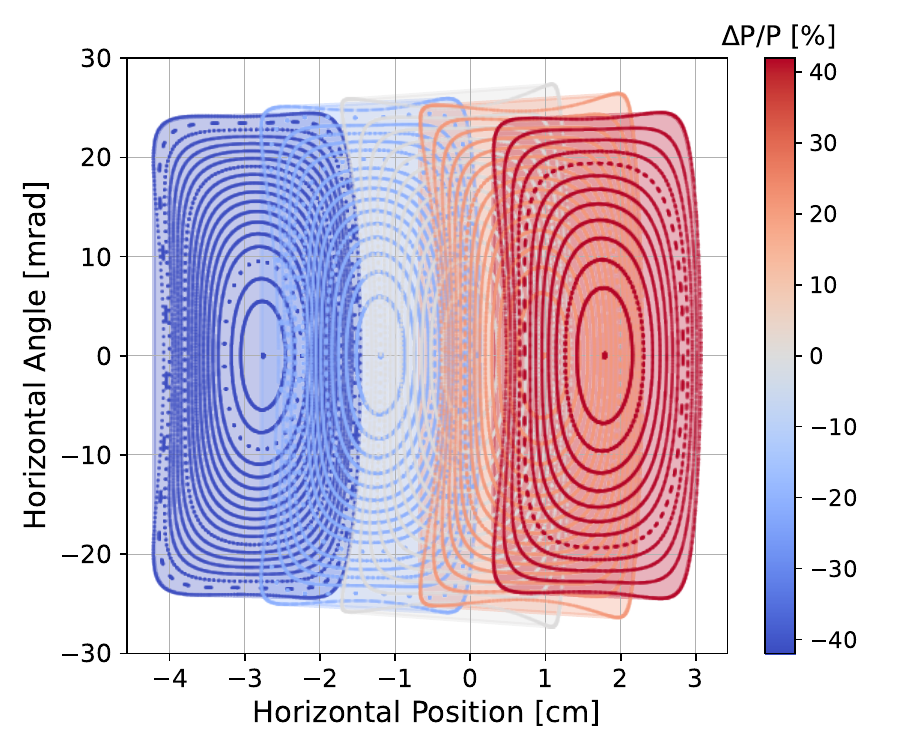}
  \caption{Stable horizontal phase space of five momenta in a single cell scaling FFA cell, measured at the start of the lattice. Closed orbit variation with momentum can be seen. The phase space is distorted at large amplitudes as the working point is close to the \nicefrac{1}{4} resonance, chosen to meet the closed-dispersion condition.}
    \label{fig: scaling_demo_phase_space}
\end{figure}

Even though the tune and CS parameters of the scaling lattice match the requirements for a closed-dispersion arc for the reference momentum, we see in Fig.~\ref{fig: scaling_disp_supp} that injecting all energies at a single point does not bring them back to the same position by the end of the arc. In Fig.~\ref{fig: scaling_demo_phase_space}, we show the phase space map for a selection of momenta, where it is clear that the origin is not inside the stable phase space volume for the full range of energies. This can be understood as an amplitude-dependent (and thus momentum-dependent) variation in the beam dynamics. This is counter to the usual operation of scaling FFAs, where phase advance is not a function of energy as particles follow closed orbits. In addition, the excursion becomes quite large in the centre of the lattice, with the lowest energy passing well beyond the expected magnet aperture. As such, this lattice is not an effective closed-dispersion arc: optimisation is required, breaking the scaling law.

\subsection{Optimisation of the Closed-Dispersion Arc}\label{subsec: disp_optimise}

An optimisation routine on the multipoles of the FFA lattice is performed, aiming to find a nonscaling FFA that comprises a viable closed-dispersion arc using similar techniques to those in \cite{fenning_high-order_2012}. In this case, there are \num{30} free parameters: each magnet has five multipoles (dipole--decapole) and there are six unique magnets, with the second half of the arc mirroring the first. During this optimisation, the fringe field profile is kept constant. 

Once an optimal solution is found for the arc optics, the methods described in Section \ref{sec: magnets} are used to attempt to design magnets that can produce the required fields and bore size. However, these magnet designs will not perfectly match those used in the arc optimisation, where differences in the fringe fields are the greatest contributing factor. As such, the optimisation is iterative: the arc and magnet routines are performed sequentially until they converge on a solution that meets the criteria for both the optics and magnet design.

We use a multi-objective genetic algorithm for the optimisation of the arc, allowing for all objectives to be achieved without imposing an a-priori weighting. We use the RNSGA-II implementation in pymoo \cite{blank_pymoo_2020}, as this allows for the addition of reference points. These reference points are required to ensure that the solutions converge towards the useful part of the solution space: for example, by adding a reference point in the solution space to keep the dispersion at the beamline midpoint just below the beampipe radius, we can ensure that solutions are physically feasible without attempting to bring the central dispersion down further at the expense of the other objectives. 

The optimisation objectives are to: 
\begin{itemize}
    \item reduce the residual dispersion at the end of the arc
    \item reduce the dispersion at the midpoint of the arc
    \item match the output CS parameters to their values at the start of the arc
\end{itemize}
For this study, tracking is performed in Zgoubi with \num{31} evenly spaced rigidities to ensure the solution is valid over the full energy range. We implement the objectives as

\begin{align*}
    f_1 = \sum & \left[ x_i (s=s_\text{end}) \right]^2, \\
    f_2 = \sum & \left[ x_i' (s=s_\text{end}) \right]^2, \\
    f_3 = \sum & \left[ x_i (s=s_\text{mid}) \right]^2, \\
    f_4 = \sum & \left[ x_i (s=s_\text{mark}) \right]^2, \\
    f_5 = \sum & \left[\beta_i(s=s_\text{end})-\beta_\text{ref}\right]^2, \\
    f_6 = \sum & \left[\alpha_i(s=s_\text{end})-\alpha_\text{ref})\right]^2, \\
\end{align*}

where $i$ indexes the \num{31} rigidities measured, and the subscripts `mid' and `end' refer to the midpoint and end of the beamline respectively. In Zgoubi, a marker was placed between the first and second cells to ensure the dispersion at the midpoint was not minimised at the expense of the dispersion at other points in the arc, as would be the case for a strong focus in a collider: this location is noted by the subscript `mark'. We set the injected beam CS parameters to $\beta_\text{ref}=$ \SI{3}{\meter} and $\alpha_\text{ref}=$ \num{0} for both the horizontal and vertical profiles. These CS parameters are assumed to be energy-independent, which can be ensured by collimation. We also set the emittance $\varepsilon$ to \SI{0.2}{\milli\meter\milli\radian}, although this has no impact on the optimisation.

\begin{figure}[]
  \centering
  \includegraphics[width=.975\linewidth]{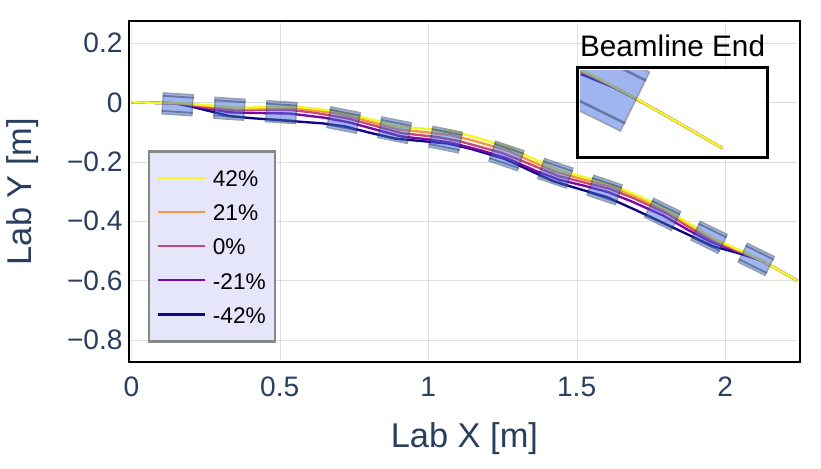}
  \caption{Trajectories in the optimised arc. By comparison with Fig.~\ref{fig: scaling_disp_supp}b), we see that the dispersion at the arc midpoint is now small enough to keep all energies within the beampipe. Dispersion at the end is also significantly reduced.}
    \label{fig: optimised_trajectories}
\end{figure}

\begin{figure}[]
  \centering
  \includegraphics[width=.9\linewidth]{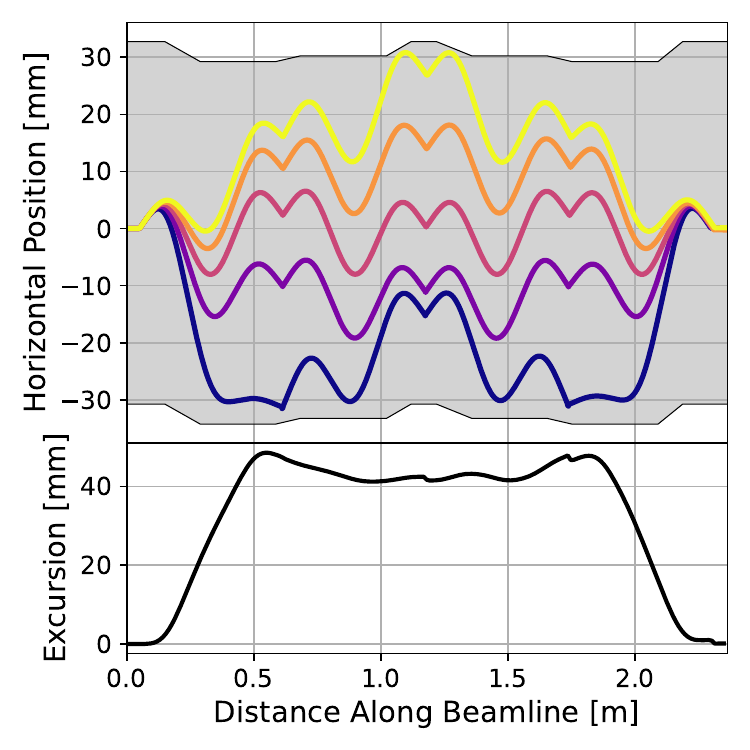}
  \caption{Orbit excursion in the optimised arc, with the approximate beampipe indicated. Colours correspond to momenta labelled in Fig.~\ref{fig: optimised_trajectories}. Discontinuities arise from rotation of the reference axis between cells: paths in real space are smooth.}
    \label{fig: optimised_excursions}
\end{figure}

In each iteration of the beamline design, the optimisation algorithm was allowed to investigate approximately \num{100000} potential solutions before termination: for example, \num{102907} configurations evaluated in the final iteration. The final optimised beamline is shown in Fig.~\ref{fig: optimised_trajectories}, with the orbit excursion shown separately in Fig.~\ref{fig: optimised_excursions}. Details of the magnet parameters are given in Table \ref{tab:magnet_details}. 

\begin{table*}
    \centering
    \begin{tabular}{lrrrrrrrr}
		\toprule
		\textbf{Magnet} & \textbf{Dipole [$10^{-1}$\si{\tesla}]} & \textbf{Quad. [$10^0$\si{\tesla/\meter}]} & \textbf{Sext. [$10^1$\si{\tesla/\meter^2}]} & \textbf{Oct. [$10^2$\si{\tesla/\meter^3}]} & \textbf{Dec. [$10^3$\si{\tesla/\meter^4}]} & \textbf{C$_0$} & \textbf{C$_1$} & \textbf{Offset [\si{\milli\meter}]} \\
		\midrule
         F\textsubscript{1} & \num{2.876}  & \num{5.331}  & \num{7.240}  & \num{3.937}  & \num{1.646}  & \num{-0.0422} & \num{4.33} & \num{0.0} \\
         D\textsubscript{1} & \num{-4.322} & \num{-11.63} & \num{-14.31} & \num{-8.358} & \num{-1.538} & \num{-0.0313} & \num{3.89} & \num{-3.5} \\
         F\textsubscript{2} & \num{2.425}  & \num{9.685}  & \num{4.867}  & \num{1.867}  & \num{32.50}  & \num{-0.0295} & \num{3.85} & \num{-3.5} \\
         F\textsubscript{3} & \num{2.828}  & \num{5.590}  & \num{8.023}  & \num{3.540}  & \num{2.264}  & \num{-0.0426} & \num{4.34} & \num{-2.5} \\
         D\textsubscript{2} & \num{-4.418} & \num{-9.760} & \num{-7.937} & \num{-8.390} & \num{-11.35} & \num{-0.0303} & \num{3.88} & \num{-2.5} \\
         F\textsubscript{4} & \num{2.861}  & \num{6.235}  & \num{6.953}  & \num{3.878}  & \num{4.314}  & \num{-0.0384} & \num{4.46} & \num{0.0} \\
        \bottomrule
    \end{tabular}
    \caption{Parameters of the magnets in the optimised TURBO arc. By offsetting the magnets horizontally, we ensure that the good field region covers the beam excursion at all points. Coefficients $C_i$ correspond to the standard Enge formula, assuming that the fringes are the same at the entrance and exit. In all cases, the magnet bore diameter $\lambda$ is \SI{7.0}{\centi\meter}.}
    \label{tab:magnet_details}
\end{table*}

In this study, the bottleneck optimiser objective was minimisation of the dispersion at the marker and midpoint. For the basic scaling lattice shown in Fig.~\ref{fig: scaling_disp_supp}, the orbit excursion range at the midpoint is \SI{11}{\centi\meter}, which is more than double the beampipe diameter. If the excursion range at this point were allowed to be so large, lattices could be found more readily with a small CS parameter mismatch and minimal residual dispersion at the end of the beamline. In our final lattice, the maximum excursion range is less than \SI{5}{\centi\meter}, and is approximately \SI{4}{\centi\meter} at the centre of the beamline: we find that any attempt to further reduce the excursion has a severe detrimental impact on the CS parameter mismatch, motivating the use of reference points in the optimisation algorithm to keep solutions in the physically useful region of the possible solution space.

\subsection{Beamline Performance}

For the TURBO project, the beamline needs to be considered from both an accelerator and clinical perspective. From an accelerator point of view the evolution of the CS parameters and the dispersion $\mathcal H$-function are of interest. Conversely, a clinical perspective focuses more on the beam size, ellipticity \cite{zhao_achieving_2021}, and spot position variation at the end of the beamline. In both cases, these vary as a function of energy, which is not usually a significant factor for clinical systems. The accelerator and medical parametrisations are related: for example, the size of the beam at the end of the beamline is given by $\sqrt{\beta\varepsilon}$.

\begin{figure}[]
  \centering
  \includegraphics[width=.9\linewidth]{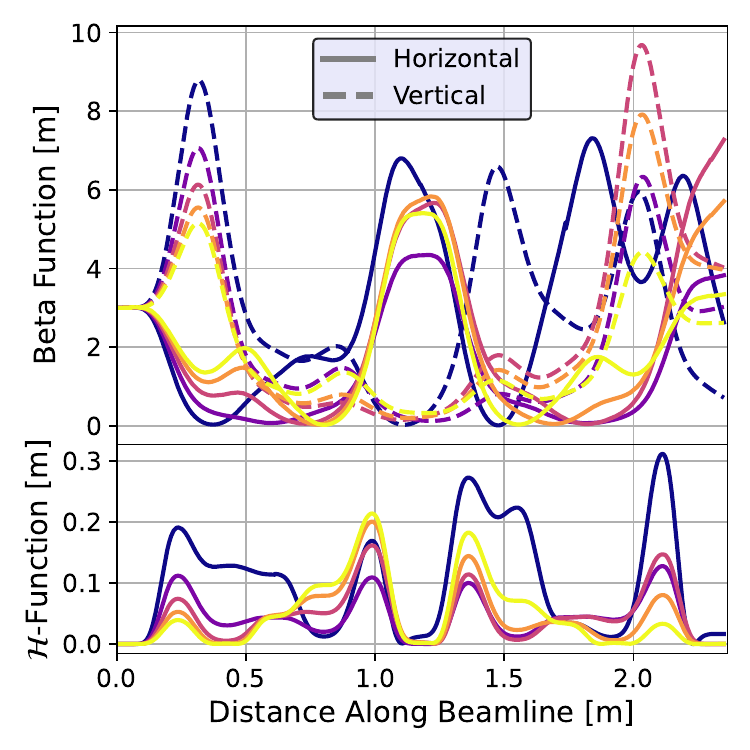}
  \caption{Evolution of the beta functions parameters and the dispersion $\mathcal H$-function. Though only five momenta are displayed, the CS parameters vary smoothly between them. Colours are the same as in Fig.~\ref{fig: optimised_trajectories}.}
    \label{fig: twiss_params}
\end{figure}

In Fig.~\ref{fig: twiss_params}, we see the CS parameters are not periodic over the length of the arc, and the centre of the arc is not a symmetry point. This can be understood as a mismatch between the initial values of $\alpha$ and $\beta$ that are implied by the beam optics, and the injected parameters which are energy-independent. The dispersion $\mathcal H$-function is not periodic for the same reason. However, the $\mathcal H$-function returns to zero at the end of the arc, indicating that the overall beamline cancels out any introduced dispersion at its end. The beta functions through the arc are kept small, although there is some distortion of the lowest energy at large amplitudes due to the nonlinear B-fields.

\begin{figure}[]
  \centering
  \includegraphics[width=.98\linewidth]{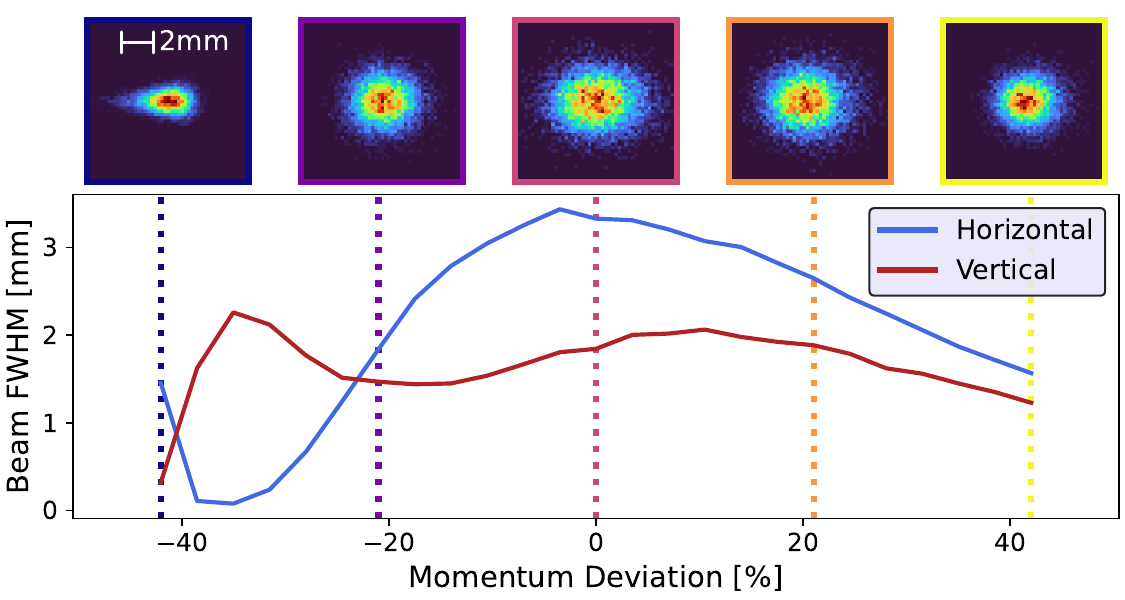}
  \caption{Final spot size as a function of momentum  deviation. Spots for five demonstrative momenta are shown ($10^4$ particles each). Beam centroid deviation is within \SI{\pm 0.2}{\milli\meter}}
    \label{fig: spot_size_and_position}
\end{figure}

The delivered beam spot quality is of paramount importance for clinical applications, and this demonstrator beamline indicates what might be achievable with a closed-dispersion arc. According to the AAPM TG 224 QA guidelines and recommendations for proton therapy \cite{arjomandy_aapm_2019}, a scanned pencil beam should have less than \SI{\pm 10}{\percent} spot size variation with the spot position deviating less than \SI{1}{\milli\meter}. Though this is not directly applicable in our case as the recommendation is relative to each individual energy, the AAPM guidelines still provide a useful indicator for clinical performance given there are currently no protocols: there are no clinical beamlines with a large energy acceptance. The variation in spot size at the end of our beamline is shown in Fig.~\ref{fig: spot_size_and_position}.

At the end of the beamline, we see that there is some variation in spot size and divergence, which is more pronounced at low momenta: as focusing strength is inversely proportional to momentum, this enhanced sensitivity is to be expected. The beam centre shift is below the AAPM recommended threshold, which is encouraging. It is expected that much of the variation with the delivered beam could be mitigated by the delivery nozzle as it plays a significant role in beam shaping, collimating, and steering for clinical treatment \cite{safai_comparison_2008}. Collimation in both planes as a function of momentum could be used to further reduce the variation in beam spot size and shape without redesigning the beamline optics \cite{geoghegan_design_2020, nelson_dosimetric_2023}: although this is standard practice for conventional X-ray radiotherapy, `dynamic collimation' is not commonplace for proton therapy. It is also not known how much scattering in air would remove differences in the beam spot shape: these points should be addressed in greater detail in future studies.

\section{Robustness of the TURBO Arc}\label{sec: robustness}

In practice, there are many errors which will impact the performance of the closed-dispersion arc. Some errors can be compensated, such as multipole errors which will be corrected as far as possible before beamline installation, and alignment errors that can be reduced by adjusting beamline elements. However, no matter how closely all the beamline parameters are controlled, it is inevitable that some discrepancy will exist between the ideal design presented in Section \ref{sec: beamline} and the beamline that is ultimately constructed.

For a closed-dispersion arc, there are two key performance indicators. The more straightforward metric is variation in beam spot position due to errors, away from the ideal case: simulations shown in Fig.~\ref{fig: spot_size_and_position} indicate that the beam spot varies by up to \SI{\pm0.2}{\milli\meter} with momentum, which we use as our threshold for acceptable beam spot position variation. To evaluate the extent of beam spot shape distortion due to errors, we use the mismatch factor $\mathcal F$ \cite{edwards_emittance_1993} between an intended set of CS parameters $(\beta_0, \alpha_0, \gamma_0)$ and those that are transported in practice $(\beta, \alpha, \gamma)$, given by

\begin{equation}\label{eqn: mismatch}
    \mathcal F = \frac{1}{2}\left(\beta_0\gamma + \beta\gamma_0-2\alpha\alpha_0\right).
\end{equation}

We use the mismatch factor $\mathcal F$ to quantify the magnitude of beam errors: if there is no mismatch (i.e. the CS parameters are exactly their intended values), $\mathcal F =$ \num{1}, otherwise it is larger. The studies here evaluate $\mathcal F -1$ at the end of the arc. We only investigate at $\mathcal F$ in the horizontal plane.

For all the error studies here, \num{500} configurations were evaluated for \num{10} different error magnitudes. Errors are assigned using a normal distribution truncated at three standard deviations to remove unrealistic outliers. The five representative energies used in the previous section were transported through each of the \num{5000} configurations. Bootstrapping \cite{belsley_bootstrap_2009} is used to find the approximate confidence intervals in all cases.

\subsection{Injected LaTeX Error: Invalid UTF-8 byte sequenceBeam Parameters}\label{subsec: injected_cs}

The beam waist ($\alpha=0$) is designed to be at the start of the closed dispersion beamline. However, fluctuations in the injected beam profile or positioning errors may shift the beam waist, impacting the output beam characteristics: this has implications in a clinical beamline or other applications. For the case of TURBO, the Pelletron output is known to have time-dependent fluctuations \cite{steinberg_characterising_2024}. As such, we study the mismatch parameter as a function of beam waist displacement to provide a baseline understanding of the relative sensitivity of the arc, to different errors.

\begin{figure}[]
  \centering
  \includegraphics[width=.95\linewidth]{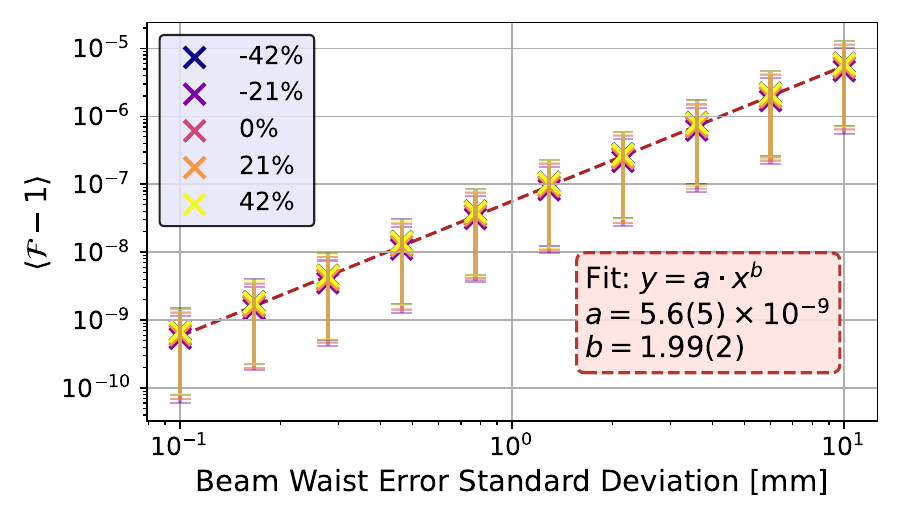}
  \caption{CS parameter mismatch as a function of beam waist offset distance, with \SI{95}{\percent} confidence intervals. $<\mathcal{F}-1>$ at the output is proportional to the $<\mathcal{F}-1>$ at the input, which goes as the square of the beam waist error.}
    \label{fig: waist_mismatch}
\end{figure}

The results of applying input beam waist displacement errors up to \SI{1}{\centi\meter} are in Fig.~\ref{fig: waist_mismatch}, which shows the output mismatch due to injection errors from a shift in the beam waist. We find that the mismatch factor at the beamline end grows with the square of the beam waist error, which is the same as the growth in $\mathcal F$ from an extra drift: this suggests that the final mismatch is proportional to the injected mismatch. As such, we conclude that the TURBO beamline design is robust against injection errors despite the highly nonlinear fields. We also see that there is not a significant energy dependence, which further suggests that the highest and lowest energies are not more sensitive to the input, despite the large excursions from the beamline centre.

One goal of the TURBO project will be to confirm that a closed-dispersion arc with nonlinear fields is as robust against injection errors as this modelling suggests. A variable matching section of four quadrupoles would be sufficient to move the beam waist between experiments. To induce more rapid beam fluctuations, a dynamic collimator would be required. To determine the beam phase space both before and after the arc, a standard pepperpot device would be sufficient. Studies are ongoing for the beam manipulation and diagnostic systems for the TURBO demonstrator beamline.

\subsection{Multipole Errors}

Multipole errors are a key driver of beam distortion and instability in all accelerators, and as such the usual standard is to keep these errors less than one part in $10^4$. As discussed in Section \ref{sec: magnets}, the TURBO beamline demonstration should achieve this, and any residual errors can be further corrected using methods such as those discussed in \cite{brooks_permanent_2020}. Other errors arise from temperature-dependent fluctuations in magnetic remanence $B_r$, or warping of the magnet mount over time. For this study, we assign errors with a standard deviation up to \SI{1}{\percent} to investigate the robustness of the arc: magnet studies in Section \ref{sec: magnets} suggest that an error up to \SI{0.1}{\percent} is expected for the TURBO beamline. Errors are applied individually to each multipole of each magnet in the arc, assuming that each error is independent. 

\begin{figure}[]
  \centering
  \includegraphics[width=.9\linewidth]{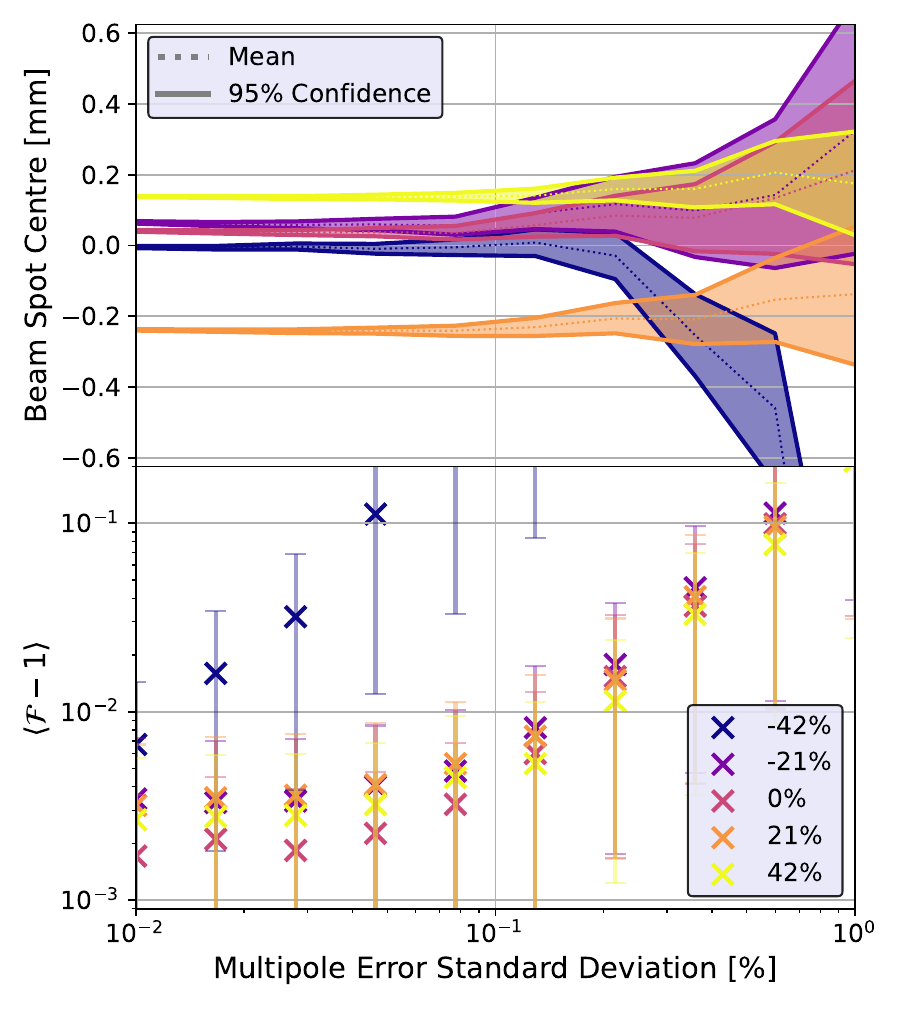}
  \caption{Beam centroid displacement and mismatch factor at the end of the beamline as a function of multipole error. In the extreme case of error values of 1\%, the lowest energy would likely impact the beampipe before being transported to the end.}
    \label{fig: multipole_mismatch}
\end{figure}

In Fig.~\ref{fig: multipole_mismatch}, we see that the closed-dispersion arc shown here is sensitive to errors, particularly for the lowest momentum. The beam spot centre experiences a small shift for most energies, with a much larger error at the lowest energy. It is interesting to note that the beam spot shift is not symmetric, even though the multipole errors are: taking the lowest energy as an extreme example, for large multipole errors the final beam spot can shift by several \si{\milli\meter}. The beam spot shift asymmetry can be understood by looking at the trajectory of the lowest momentum in Fig.~\ref{fig: optimised_excursions}: due to the large negative excursion, multipole errors tend to push the beam further out from the reference axis. At the \SI{0.1}{\percent} multipole error that can be achieved for the TURBO arc, the beam spot centre shift in Fig.~\ref{fig: multipole_mismatch} is smaller than the intrinsic offsets from the beamline, meeting our design requirements. 

The mismatch factor error is several orders of magnitude larger than we predict for beam waist errors. At points in the arc where the CS beta function is small, we expect the multipole error sensitivity to be enhanced. In particular, the point where $\beta_x$ is small near the start of the arc (around \SI{0.3}{\meter} in Fig.~\ref{fig: twiss_params}) will be particularly susceptible, as even small errors here will be compounded and amplified by the end of the arc. Once again the lowest energy appears particularly sensitive, corresponding to it having the smallest value of $\beta_x$ at this point. 

From these results, magnet multipole inaccuracies are shown to dominate the delivered beam  errors, so it is important to consider possible mitigations. One option is to mount each magnet on a translating stand with several degrees of freedom, providing some correction to multipole errors through harmonic feed-down. Although this correction technique proved effective for orbit corrections in CBETA, it would drive up the overall cost of the beamline. Alternatively, the multipole correction method applied before installation could be tuned to provide extra compensation to lower energies, which may reduce the mismatch seen here for the lowest momentum at the expense of increased mismatch in other cases.

\subsection{Alignment Errors}

\begin{figure}[]
  \centering
  \includegraphics[width=.9\linewidth]{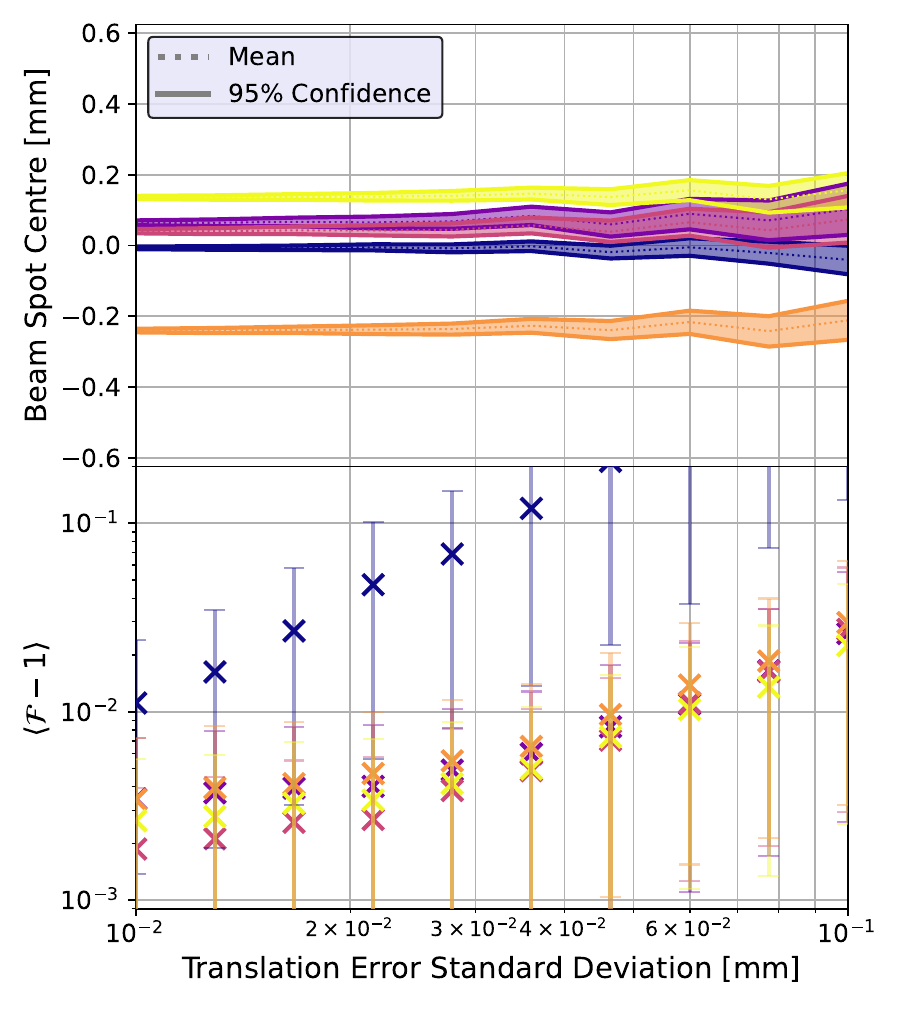}
  \caption{Beam spot and mismatch error due to magnet translation in the horizontal plane. The vertical axes and colour scale are identical to Figure \ref{fig: multipole_mismatch}.}
    \label{fig: translation_mismatch}
\end{figure}

Magnet misalignments are another common source of error in accelerators. In many cases, sophisticated laser systems and girder assembly enable alignment to sub-\SI{0.1}{\milli\meter} precision \cite{vikas_review_2021}. For the TURBO beamline, vibrations in the room and misalignments on the girder may be more difficult to mitigate. In addition, the relative positions of the beampipe, the magnets, and the Pelletron itself may shift over the course of a day as the building itself moves by a small amount. With proper isolation of the equipment from the surroundings and careful maintenance, it is expected that most misalignments can be kept below \SI{100}{\micro\meter}: the impacts of these errors on the beamline outputs are shown in Fig.~\ref{fig: translation_mismatch}. To facilitate the simulation of misalignment errors, we use the Zgoubi \texttt{DRM} parameter, which enables offsets in the horizontal plane. 

Qualitatively, the impact of misalignments appears similar to the multipole errors in Fig.~\ref{fig: multipole_mismatch}, with a similar shape to the mismatch factor curve and the highest sensitivity for the lowest momentum: this similarity is to be expected, as the highly nonlinear magnetic fields convert any position offset to a feed-down multipole error. The performance of the arc is not significantly degraded, even where errors get up to around \SI{0.1}{\milli\meter}, which suggests that the feed-down multipole error will not be the limiting factor for beamline utility. Although only horizontal misalignments have been analysed, it is expected that full 3D translation and rotation would not be significantly different.

\section{Discussion}

We have devised a method of designing closed-dispersion arcs  using FFA optics, enabling transport of a large range of momenta without adjusting magnetic fields. This is achieved by using multipoles up to decapole order to provide sufficient beam steering and focusing over the full momentum range. Applying this idea to a scaled-down prototype for the TURBO project, we have designed a beamline  to transport \num{0.5}--\SI{3.0}{\mega\electronvolt} proton beams (\SI{\pm 42}{\percent} momentum acceptance) produced by the Pelletron at the University of Melbourne. In addition to a basic optics design, we have produced a method to design suitable Halbach arrays for the arc using low-cost commercial permanent magnets, and error studies have been performed to better understand the beamline performance under realistic operating conditions.

Permanent magnets will be used for the TURBO beamline as field ramping is not required, and they can produce fields in excess of \SI{0.5}{\tesla} over a large aperture. Previous FFA designs with permanent magnets required custom wedges of magnetic material to make a Halbach array with the desired field, however these wedges are only suitable for a single use application and may be expensive to manufacture. For this project, we have found a method to design Halbach arrays using a single block shape with multipole errors below one part in $10^4$, by approximating the required magnetisation that would be used in the ideal case and optimising the positions of the individual blocks. By using a single type of magnet block, we hope to use commercially available permanent magnets inserted in a simple 3D printed or aluminum-milled housing, which should significantly reduce costs, and allows for rapid prototyping if the magnets must be redesigned. The magnetic material can also be reused for future projects, reducing waste and overall costs over the beamline lifetime.

For the beamline itself, we have presented a generic method for designing closed-dispersion arcs with Fixed Field Accelerator optics that is able to transport a large range of momenta, as applied to the TURBO project. Our method begins with an approximately scaling FFA triplet cell with a phase advance of \nicefrac{$\pi$}{2}, with four cells to introduce and remove dispersion. We show that a scaling FFA is not sufficient to remove dispersion over such a large range of momenta, and requires a large beam excursion at the midpoint of the arc. The symmetry of the arc requires that the second half mirror the first: as we go up to decapole order multipoles, there are 30 optimiser variables. To solve this, a genetic algorithm with reference points is used to explore the solution space. 

This algorithm produced a suitable lattice for the TURBO beamline, meeting our key design requirements with a maximum field below \SI{0.6}{\tesla} and an excursion range below \SI{50}{\milli\meter} while fitting inside our required beamline geometry. In the optimisation process, we found that there is a trade-off between matching the CS parameters, and returning all trajectories back to the origin. In this case, there may simply not be sufficient degrees of freedom to match all energies perfectly: in the future, a design going beyond decapole terms could be considered to give better matching. In addition, the number of magnets in the arc could be increased, either by reducing the drift lengths or by increasing the total length of the arc. However, both of these potential solutions raise further challenges, such as an increased B-field overlap between adjacent magnets, stricter spacial constraints, and increased costs. Despite the observed optics mismatch, the variation in beam spot size and shape as a function of energy is acceptable, and may be further mitigated using methods such as dynamic collimation if required.

Using a scaling FFA as the basis for a closed-dispersion arc provides a good starting point for further optimisation, as it ensures stable optics exist along closed orbits for all momenta. However, there are disadvantages to this choice of initial optics. The scaling law determines the dispersion in the initial lattice, increasing the difficulty of later optimisation. In addition, our optimisation does not allow the polarity of any magnets to change, so the reverse bending from the `D' magnets remains significant. As such, the overall arc is larger than would be necessary for a lattice with minimal reverse bending, and the magnetic fields required throughout the arc are larger than would be otherwise required. A future study could investigate the design of a closed-dispersion arc that does not use a scaling optics as the initial solution, however it may be difficult to find a suitable optics given the large possible solution space. 

It should be noted that the beamline design presented here does not use the realistic model created with the method described in Section \ref{sec: magnets}, as the design iterations did not converge. The limiting factor is the large fringe field extent: the fringe fields vary significantly with each magnet iteration, as they are strongly dependent on the positioning of the magnet blocks. In addition, we find that attempts to reduce the residual dispersion further than is achieved here are generally not successful, due to the contributions of overlapping fringe fields counteracting one another. This could be mitigated by reducing the fringe field extent and variability, either by reducing the magnet aperture or by the introduction of field clamps \cite{hubner_design_1970, schermer_tuning_1987}. These options should be investigated to improve the performance of the final TURBO beamline design.

Our initial error study of the TURBO lattice has suggested that closed-dispersion arcs are robust against realistic misalignments and multipole errors, which can be managed by standard mitigation techniques. The highly nonlinear fields and large beam excursions make all energies susceptible to these errors, although the lowest was found to be particularly sensitive for the lattice shown here as it spends the longest time at the largest excursion. This may be countered by increasing the relative weighting of low rigidity beams during the optimisation process. Further work is required to fully investigate the effectiveness of error mitigation techniques, but the initial tolerance study shown here suggests that it will be possible to ensure good beam delivery for all momenta.

\section{Conclusions}\label{sec: conclusions}

To act as a closed-dispersion arc over a large range of momenta, the TURBO beamline will be a nonlinear FFA lattice suitable for the beam available at the University of Melbourne Pelletron lab, transporting proton beams between \num{0.5}--\SI{3.0}{\mega\electronvolt}. This momentum range is equivalent to \num{46}--\SI{250}{\mega\electronvolt} protons when scaled up to the clinical range. By covering a greater momentum acceptance than is required for standard proton therapy, TURBO will also demonstrate possible extensions to proton CT and ion therapy. The arc has been designed such that all energies are overlayed at the start and the end, without significant distortion of the output beam spot. Initial magnet designs have also been completed, using commercially available blocks of permanent magnet material to reproduce the required magnetic fields. This study is supplemented by analysis of the errors that may arise from realistic magnet misalignments, finding that the beamline is robust in realistic conditions, however further optimisation may be required to ensure good performance.

The ultimate goal of the TURBO project is to design and demonstrate a large energy acceptance for the energy range relevant to charged particle therapy. The low energy demonstrator discussed here has explored many of the challenges that will be faced by a clinical closed-dispersion arc, such as ensuring the beam excursion is sufficiently small and maintaining good beam quality in spite of errors. The magnets shown here would not be well-suited to a higher energy, as they would not be able to produce sufficient field to keep the beamline compact: instead, it is expected that superconducting curved canted cosine theta magnets would be able to provide the strong beam steering and focusing required \cite{caspi_cantedcosinetheta_2014, wan_alternating-gradient_2015, zhao_design_2020}. In addition, these magnets are able to produce a highly nonlinear magnetic field over a large aperture, a key requirement shown in this work. Investigations will be required to ensure the field quality is sufficient, in particular due to the fringe fields associated with these magnets, and this will form part of the full-scale beamline design at a later stage.

This study is a first investigation of a large energy acceptance arc for the proof-of-principle TURBO beamline, and there are many avenues for further investigation. This work also opens up the possibility of applying a large energy acceptance closed-dispersion arc as an insertion as part of a ring. By providing a design method for feasible permanent magnets, an initial lattice design, and an analysis of likely machine errors, we have shown that a closed-dispersion arc at scaled-down energies will be able to demonstrate viability of large momentum acceptance beamlines for charged particle therapy.

\section{Acknowledgements}\label{sec: acknowledgements}
The authors are grateful for the support of Hannah Norman and Paarangat Pushkarna, whose feedback, comments, and advice were of great assistance. Funding for the contributors on this project is through the Universities of Manchester and Melbourne Cookson Scholarship, the Laby Foundation, and ANSTO.

\bibliographystyle{apsrev}
\bibliography{references}

\begin{thebibliography}{86}
\expandafter\ifx\csname natexlab\endcsname\relax\def\natexlab#1{#1}\fi
\expandafter\ifx\csname bibnamefont\endcsname\relax
  \def\bibnamefont#1{#1}\fi
\expandafter\ifx\csname bibfnamefont\endcsname\relax
  \def\bibfnamefont#1{#1}\fi
\expandafter\ifx\csname citenamefont\endcsname\relax
  \def\citenamefont#1{#1}\fi
\expandafter\ifx\csname url\endcsname\relax
  \def\url#1{\texttt{#1}}\fi
\expandafter\ifx\csname urlprefix\endcsname\relax\def\urlprefix{URL }\fi
\providecommand{\bibinfo}[2]{#2}
\providecommand{\eprint}[2][]{\url{#2}}

\bibitem[{noa()}]{noauthor_particle_nodate}
\emph{\bibinfo{title}{The {Particle} {Therapy} {Co}-{Operative} {Group}}}, \urlprefix\url{https://www.ptcog.site/index.php}.

\bibitem[{\citenamefont{Olsen et~al.}(2007)\citenamefont{Olsen, Bruland, Frykholm, and Norderhaug}}]{olsen_proton_2007}
\bibinfo{author}{\bibfnamefont{D.~R.} \bibnamefont{Olsen}}, \bibinfo{author}{\bibfnamefont{Ã.~S.} \bibnamefont{Bruland}}, \bibinfo{author}{\bibfnamefont{G.}~\bibnamefont{Frykholm}}, \bibnamefont{and} \bibinfo{author}{\bibfnamefont{I.~N.} \bibnamefont{Norderhaug}}, \bibinfo{journal}{Radiotherapy and Oncology} \textbf{\bibinfo{volume}{83}}, \bibinfo{pages}{123} (\bibinfo{year}{2007}), ISSN \bibinfo{issn}{01678140}, \urlprefix\url{https://linkinghub.elsevier.com/retrieve/pii/S016781400700093X}.

\bibitem[{\citenamefont{Burnet et~al.}(2022)\citenamefont{Burnet, Mee, Gaito, Kirkby, Aitkenhead, Anandadas, Aznar, Barraclough, Borst, Charlwood et~al.}}]{burnet_estimating_2022}
\bibinfo{author}{\bibfnamefont{N.~G.} \bibnamefont{Burnet}}, \bibinfo{author}{\bibfnamefont{T.}~\bibnamefont{Mee}}, \bibinfo{author}{\bibfnamefont{S.}~\bibnamefont{Gaito}}, \bibinfo{author}{\bibfnamefont{N.~F.} \bibnamefont{Kirkby}}, \bibinfo{author}{\bibfnamefont{A.~H.} \bibnamefont{Aitkenhead}}, \bibinfo{author}{\bibfnamefont{C.~N.} \bibnamefont{Anandadas}}, \bibinfo{author}{\bibfnamefont{M.~C.} \bibnamefont{Aznar}}, \bibinfo{author}{\bibfnamefont{L.~H.} \bibnamefont{Barraclough}}, \bibinfo{author}{\bibfnamefont{G.}~\bibnamefont{Borst}}, \bibinfo{author}{\bibfnamefont{F.~C.} \bibnamefont{Charlwood}}, \bibnamefont{et~al.}, \bibinfo{journal}{The British Journal of Radiology} \textbf{\bibinfo{volume}{95}}, \bibinfo{pages}{20211175} (\bibinfo{year}{2022}), ISSN \bibinfo{issn}{0007-1285, 1748-880X}, \urlprefix\url{https://academic.oup.com/bjr/article/7477394}.

\bibitem[{\citenamefont{Mohan}(2022)}]{mohan_review_2022}
\bibinfo{author}{\bibfnamefont{R.}~\bibnamefont{Mohan}}, \bibinfo{journal}{Precision Radiation Oncology} \textbf{\bibinfo{volume}{6}}, \bibinfo{pages}{164} (\bibinfo{year}{2022}), ISSN \bibinfo{issn}{2398-7324, 2398-7324}, \urlprefix\url{https://onlinelibrary.wiley.com/doi/10.1002/pro6.1149}.

\bibitem[{\citenamefont{Mohan and Grosshans}(2017)}]{mohan_proton_2017}
\bibinfo{author}{\bibfnamefont{R.}~\bibnamefont{Mohan}} \bibnamefont{and} \bibinfo{author}{\bibfnamefont{D.}~\bibnamefont{Grosshans}}, \bibinfo{journal}{Advanced Drug Delivery Reviews} \textbf{\bibinfo{volume}{109}}, \bibinfo{pages}{26} (\bibinfo{year}{2017}), ISSN \bibinfo{issn}{0169409X}, \urlprefix\url{https://linkinghub.elsevier.com/retrieve/pii/S0169409X16303192}.

\bibitem[{\citenamefont{Paganetti}(2016)}]{paganetti_proton_2016}
\bibinfo{author}{\bibfnamefont{H.}~\bibnamefont{Paganetti}}, \emph{\bibinfo{title}{Proton {Beam} {Therapy}}} (\bibinfo{publisher}{IOP Publishing}, \bibinfo{year}{2016}), ISBN \bibinfo{isbn}{978-0-7503-1370-4}, \urlprefix\url{http://iopscience.iop.org/book/978-0-7503-1370-4}.

\bibitem[{\citenamefont{Yamada}(1995)}]{yamada_commissioning_1995}
\bibinfo{author}{\bibfnamefont{S.}~\bibnamefont{Yamada}}, \bibinfo{journal}{PAC 1995} pp. \bibinfo{pages}{9--13 vol.1} (\bibinfo{year}{1995}).

\bibitem[{\citenamefont{Suzuki et~al.}(2011)\citenamefont{Suzuki, Gillin, Sahoo, Zhu, Lee, and Lippy}}]{suzuki_quantitative_2011}
\bibinfo{author}{\bibfnamefont{K.}~\bibnamefont{Suzuki}}, \bibinfo{author}{\bibfnamefont{M.~T.} \bibnamefont{Gillin}}, \bibinfo{author}{\bibfnamefont{N.}~\bibnamefont{Sahoo}}, \bibinfo{author}{\bibfnamefont{X.~R.} \bibnamefont{Zhu}}, \bibinfo{author}{\bibfnamefont{A.~K.} \bibnamefont{Lee}}, \bibnamefont{and} \bibinfo{author}{\bibfnamefont{D.}~\bibnamefont{Lippy}}, \bibinfo{journal}{Medical Physics} \textbf{\bibinfo{volume}{38}}, \bibinfo{pages}{4329} (\bibinfo{year}{2011}), ISSN \bibinfo{issn}{0094-2405, 2473-4209}, \urlprefix\url{https://aapm.onlinelibrary.wiley.com/doi/10.1118/1.3604153}.

\bibitem[{\citenamefont{Shen et~al.}(2017)\citenamefont{Shen, Tryggestad, Younkin, Keole, Furutani, Kang, Herman, and Bues}}]{shen_technical_2017}
\bibinfo{author}{\bibfnamefont{J.}~\bibnamefont{Shen}}, \bibinfo{author}{\bibfnamefont{E.}~\bibnamefont{Tryggestad}}, \bibinfo{author}{\bibfnamefont{J.~E.} \bibnamefont{Younkin}}, \bibinfo{author}{\bibfnamefont{S.~R.} \bibnamefont{Keole}}, \bibinfo{author}{\bibfnamefont{K.~M.} \bibnamefont{Furutani}}, \bibinfo{author}{\bibfnamefont{Y.}~\bibnamefont{Kang}}, \bibinfo{author}{\bibfnamefont{M.~G.} \bibnamefont{Herman}}, \bibnamefont{and} \bibinfo{author}{\bibfnamefont{M.}~\bibnamefont{Bues}}, \bibinfo{journal}{Medical Physics} \textbf{\bibinfo{volume}{44}}, \bibinfo{pages}{5081} (\bibinfo{year}{2017}), ISSN \bibinfo{issn}{2473-4209}.

\bibitem[{\citenamefont{Paganetti et~al.}(2021)\citenamefont{Paganetti, Beltran, Both, Dong, Flanz, Furutani, Grassberger, Grosshans, Knopf, Langendijk et~al.}}]{paganetti_roadmap_2021}
\bibinfo{author}{\bibfnamefont{H.}~\bibnamefont{Paganetti}}, \bibinfo{author}{\bibfnamefont{C.}~\bibnamefont{Beltran}}, \bibinfo{author}{\bibfnamefont{S.}~\bibnamefont{Both}}, \bibinfo{author}{\bibfnamefont{L.}~\bibnamefont{Dong}}, \bibinfo{author}{\bibfnamefont{J.}~\bibnamefont{Flanz}}, \bibinfo{author}{\bibfnamefont{K.}~\bibnamefont{Furutani}}, \bibinfo{author}{\bibfnamefont{C.}~\bibnamefont{Grassberger}}, \bibinfo{author}{\bibfnamefont{D.~R.} \bibnamefont{Grosshans}}, \bibinfo{author}{\bibfnamefont{A.-C.} \bibnamefont{Knopf}}, \bibinfo{author}{\bibfnamefont{J.~A.} \bibnamefont{Langendijk}}, \bibnamefont{et~al.}, \bibinfo{journal}{Physics in Medicine \& Biology} \textbf{\bibinfo{volume}{66}}, \bibinfo{pages}{05RM01} (\bibinfo{year}{2021}), ISSN \bibinfo{issn}{0031-9155, 1361-6560}, \urlprefix\url{https://iopscience.iop.org/article/10.1088/1361-6560/abcd16}.

\bibitem[{\citenamefont{Yap et~al.}(2021)\citenamefont{Yap, De~Franco, and Sheehy}}]{yap_future_2021}
\bibinfo{author}{\bibfnamefont{J.}~\bibnamefont{Yap}}, \bibinfo{author}{\bibfnamefont{A.}~\bibnamefont{De~Franco}}, \bibnamefont{and} \bibinfo{author}{\bibfnamefont{S.}~\bibnamefont{Sheehy}}, \bibinfo{journal}{Frontiers in Oncology} \textbf{\bibinfo{volume}{11}}, \bibinfo{pages}{780025} (\bibinfo{year}{2021}), ISSN \bibinfo{issn}{2234-943X}, \urlprefix\url{https://www.frontiersin.org/articles/10.3389/fonc.2021.780025/full}.

\bibitem[{\citenamefont{Grassberger et~al.}(2013)\citenamefont{Grassberger, Dowdell, Lomax, Sharp, Shackleford, Choi, Willers, and Paganetti}}]{grassberger_motion_2013}
\bibinfo{author}{\bibfnamefont{C.}~\bibnamefont{Grassberger}}, \bibinfo{author}{\bibfnamefont{S.}~\bibnamefont{Dowdell}}, \bibinfo{author}{\bibfnamefont{A.~J.} \bibnamefont{Lomax}}, \bibinfo{author}{\bibfnamefont{G.}~\bibnamefont{Sharp}}, \bibinfo{author}{\bibfnamefont{J.}~\bibnamefont{Shackleford}}, \bibinfo{author}{\bibfnamefont{N.}~\bibnamefont{Choi}}, \bibinfo{author}{\bibfnamefont{H.}~\bibnamefont{Willers}}, \bibnamefont{and} \bibinfo{author}{\bibfnamefont{H.}~\bibnamefont{Paganetti}}, \bibinfo{journal}{International Journal of Radiation Oncology} \textbf{\bibinfo{volume}{86}} (\bibinfo{year}{2013}).

\bibitem[{\citenamefont{Engelsman et~al.}(2013)\citenamefont{Engelsman, Schwarz, and Dong}}]{engelsman_physics_2013}
\bibinfo{author}{\bibfnamefont{M.}~\bibnamefont{Engelsman}}, \bibinfo{author}{\bibfnamefont{M.}~\bibnamefont{Schwarz}}, \bibnamefont{and} \bibinfo{author}{\bibfnamefont{L.}~\bibnamefont{Dong}}, \bibinfo{journal}{Seminars in Radiation Oncology} \textbf{\bibinfo{volume}{23}}, \bibinfo{pages}{88} (\bibinfo{year}{2013}).

\bibitem[{\citenamefont{Bert and Durante}(2011)}]{bert_motion_2011}
\bibinfo{author}{\bibfnamefont{C.}~\bibnamefont{Bert}} \bibnamefont{and} \bibinfo{author}{\bibfnamefont{M.}~\bibnamefont{Durante}}, \bibinfo{journal}{Physics in Medicine and Biology} \textbf{\bibinfo{volume}{56}}, \bibinfo{pages}{R113} (\bibinfo{year}{2011}), ISSN \bibinfo{issn}{0031-9155, 1361-6560}, \urlprefix\url{https://iopscience.iop.org/article/10.1088/0031-9155/56/16/R01}.

\bibitem[{\citenamefont{Mohan et~al.}(2017)\citenamefont{Mohan, Das, and Ling}}]{mohan_empowering_2017}
\bibinfo{author}{\bibfnamefont{R.}~\bibnamefont{Mohan}}, \bibinfo{author}{\bibfnamefont{I.~J.} \bibnamefont{Das}}, \bibnamefont{and} \bibinfo{author}{\bibfnamefont{C.~C.} \bibnamefont{Ling}}, \bibinfo{journal}{International Journal of Radiation Oncology} \textbf{\bibinfo{volume}{99}}, \bibinfo{pages}{304} (\bibinfo{year}{2017}), ISSN \bibinfo{issn}{03603016}, \urlprefix\url{https://linkinghub.elsevier.com/retrieve/pii/S0360301617309021}.

\bibitem[{\citenamefont{Diffenderfer et~al.}(2022)\citenamefont{Diffenderfer, Sørensen, Mazal, and Carlson}}]{diffenderfer_current_2022}
\bibinfo{author}{\bibfnamefont{E.~S.} \bibnamefont{Diffenderfer}}, \bibinfo{author}{\bibfnamefont{B.~S.} \bibnamefont{Sørensen}}, \bibinfo{author}{\bibfnamefont{A.}~\bibnamefont{Mazal}}, \bibnamefont{and} \bibinfo{author}{\bibfnamefont{D.~J.} \bibnamefont{Carlson}}, \bibinfo{journal}{Medical Physics} \textbf{\bibinfo{volume}{49}}, \bibinfo{pages}{2039} (\bibinfo{year}{2022}), ISSN \bibinfo{issn}{2473-4209}, \bibinfo{note}{\_eprint: https://onlinelibrary.wiley.com/doi/pdf/10.1002/mp.15276}, \urlprefix\url{https://onlinelibrary.wiley.com/doi/abs/10.1002/mp.15276}.

\bibitem[{\citenamefont{Jolly et~al.}(2020)\citenamefont{Jolly, Owen, Schippers, and Welsch}}]{jolly_technical_2020}
\bibinfo{author}{\bibfnamefont{S.}~\bibnamefont{Jolly}}, \bibinfo{author}{\bibfnamefont{H.~L.} \bibnamefont{Owen}}, \bibinfo{author}{\bibfnamefont{M.}~\bibnamefont{Schippers}}, \bibnamefont{and} \bibinfo{author}{\bibfnamefont{C.}~\bibnamefont{Welsch}}, \bibinfo{journal}{Physica Medica} \textbf{\bibinfo{volume}{78}}, \bibinfo{pages}{71} (\bibinfo{year}{2020}).

\bibitem[{\citenamefont{Verhaegen et~al.}(2021)\citenamefont{Verhaegen, Wanders, Wolfs, and Eekers}}]{verhaegen_considerations_2021}
\bibinfo{author}{\bibfnamefont{F.}~\bibnamefont{Verhaegen}}, \bibinfo{author}{\bibfnamefont{R.-G.} \bibnamefont{Wanders}}, \bibinfo{author}{\bibfnamefont{C.}~\bibnamefont{Wolfs}}, \bibnamefont{and} \bibinfo{author}{\bibfnamefont{D.}~\bibnamefont{Eekers}}, \bibinfo{journal}{Physics in Medicine \& Biology} \textbf{\bibinfo{volume}{66}}, \bibinfo{pages}{06NT01} (\bibinfo{year}{2021}), ISSN \bibinfo{issn}{0031-9155, 1361-6560}, \urlprefix\url{https://iopscience.iop.org/article/10.1088/1361-6560/abe55a}.

\bibitem[{\citenamefont{{Badano} et~al.}(2000)\citenamefont{{Badano}, {Benedikt}, {Bryant}, {Crescenti}, {Holy}, {Maier}, {Pullia}, {Reimoser}, and {Rossi}}}]{badano_proton-ion_2000}
\bibinfo{author}{\bibnamefont{{Badano}}}, \bibinfo{author}{\bibnamefont{{Benedikt}}}, \bibinfo{author}{\bibnamefont{{Bryant}}}, \bibinfo{author}{\bibnamefont{{Crescenti}}}, \bibinfo{author}{\bibnamefont{{Holy}}}, \bibinfo{author}{\bibnamefont{{Maier}}}, \bibinfo{author}{\bibnamefont{{Pullia}}}, \bibinfo{author}{\bibnamefont{{Reimoser}}}, \bibnamefont{and} \bibinfo{author}{\bibnamefont{{Rossi}}}, \emph{\bibinfo{title}{{PROTON}-{ION} {MEDICAL} {MACHINE} {STUDY} ({PIMMS}) {PART} {II}}} (\bibinfo{year}{2000}), \urlprefix\url{http://cds.cern.ch/record/449577/files/ps-2000-007.pdf?version=1}.

\bibitem[{\citenamefont{Schulte et~al.}(2004)\citenamefont{Schulte, Bashkirov, {Tianfang Li}, {Zhengrong Liang}, Mueller, Heimann, Johnson, Keeney, Sadrozinski, Seiden et~al.}}]{schulte_conceptual_2004}
\bibinfo{author}{\bibfnamefont{R.}~\bibnamefont{Schulte}}, \bibinfo{author}{\bibfnamefont{V.}~\bibnamefont{Bashkirov}}, \bibinfo{author}{\bibnamefont{{Tianfang Li}}}, \bibinfo{author}{\bibnamefont{{Zhengrong Liang}}}, \bibinfo{author}{\bibfnamefont{K.}~\bibnamefont{Mueller}}, \bibinfo{author}{\bibfnamefont{J.}~\bibnamefont{Heimann}}, \bibinfo{author}{\bibfnamefont{L.}~\bibnamefont{Johnson}}, \bibinfo{author}{\bibfnamefont{B.}~\bibnamefont{Keeney}}, \bibinfo{author}{\bibfnamefont{H.-W.} \bibnamefont{Sadrozinski}}, \bibinfo{author}{\bibfnamefont{A.}~\bibnamefont{Seiden}}, \bibnamefont{et~al.}, \bibinfo{journal}{IEEE Transactions on Nuclear Science} \textbf{\bibinfo{volume}{51}}, \bibinfo{pages}{866} (\bibinfo{year}{2004}), ISSN \bibinfo{issn}{0018-9499}, \urlprefix\url{http://ieeexplore.ieee.org/document/1311983/}.

\bibitem[{\citenamefont{Bartnik et~al.}(2020)\citenamefont{Bartnik, Banerjee, Burke, Crittenden, Deitrick, Dobbins, Gulliford, Hoffstaetter, Li, Lou et~al.}}]{bartnik_cbeta_2020}
\bibinfo{author}{\bibfnamefont{A.}~\bibnamefont{Bartnik}}, \bibinfo{author}{\bibfnamefont{N.}~\bibnamefont{Banerjee}}, \bibinfo{author}{\bibfnamefont{D.}~\bibnamefont{Burke}}, \bibinfo{author}{\bibfnamefont{J.}~\bibnamefont{Crittenden}}, \bibinfo{author}{\bibfnamefont{K.}~\bibnamefont{Deitrick}}, \bibinfo{author}{\bibfnamefont{J.}~\bibnamefont{Dobbins}}, \bibinfo{author}{\bibfnamefont{C.}~\bibnamefont{Gulliford}}, \bibinfo{author}{\bibfnamefont{G.}~\bibnamefont{Hoffstaetter}}, \bibinfo{author}{\bibfnamefont{Y.}~\bibnamefont{Li}}, \bibinfo{author}{\bibfnamefont{W.}~\bibnamefont{Lou}}, \bibnamefont{et~al.}, \bibinfo{journal}{Physical Review Letters} \textbf{\bibinfo{volume}{125}}, \bibinfo{pages}{044803} (\bibinfo{year}{2020}), ISSN \bibinfo{issn}{0031-9007, 1079-7114}, \urlprefix\url{https://link.aps.org/doi/10.1103/PhysRevLett.125.044803}.

\bibitem[{\citenamefont{Aschenauer et~al.}(2014)\citenamefont{Aschenauer, Baker, Bazilevsky, Boyle, Belomestnykh, Ben-Zvi, Brooks, Brutus, Burton, Fazio et~al.}}]{aschenauer_erhic_2014}
\bibinfo{author}{\bibfnamefont{E.~C.} \bibnamefont{Aschenauer}}, \bibinfo{author}{\bibfnamefont{M.~D.} \bibnamefont{Baker}}, \bibinfo{author}{\bibfnamefont{A.}~\bibnamefont{Bazilevsky}}, \bibinfo{author}{\bibfnamefont{K.}~\bibnamefont{Boyle}}, \bibinfo{author}{\bibfnamefont{S.}~\bibnamefont{Belomestnykh}}, \bibinfo{author}{\bibfnamefont{I.}~\bibnamefont{Ben-Zvi}}, \bibinfo{author}{\bibfnamefont{S.}~\bibnamefont{Brooks}}, \bibinfo{author}{\bibfnamefont{C.}~\bibnamefont{Brutus}}, \bibinfo{author}{\bibfnamefont{T.}~\bibnamefont{Burton}}, \bibinfo{author}{\bibfnamefont{S.}~\bibnamefont{Fazio}}, \bibnamefont{et~al.} (\bibinfo{year}{2014}), \bibinfo{note}{publisher: arXiv Version Number: 2}, \urlprefix\url{https://arxiv.org/abs/1409.1633}.

\bibitem[{\citenamefont{Symon et~al.}(1956)\citenamefont{Symon, Kerst, Jones, Laslett, and Terwilliger}}]{symon_fixed-field_1956}
\bibinfo{author}{\bibfnamefont{K.~R.} \bibnamefont{Symon}}, \bibinfo{author}{\bibfnamefont{D.~W.} \bibnamefont{Kerst}}, \bibinfo{author}{\bibfnamefont{L.~W.} \bibnamefont{Jones}}, \bibinfo{author}{\bibfnamefont{L.~J.} \bibnamefont{Laslett}}, \bibnamefont{and} \bibinfo{author}{\bibfnamefont{K.~M.} \bibnamefont{Terwilliger}}, \bibinfo{journal}{Physical Review} \textbf{\bibinfo{volume}{103}}, \bibinfo{pages}{1837} (\bibinfo{year}{1956}), ISSN \bibinfo{issn}{0031-899X}, \urlprefix\url{https://link.aps.org/doi/10.1103/PhysRev.103.1837}.

\bibitem[{\citenamefont{Sessler et~al.}(2010)\citenamefont{Sessler, Mills, Jones, Symon, and Young}}]{sessler_innovation_2010}
\bibinfo{author}{\bibfnamefont{A.}~\bibnamefont{Sessler}}, \bibinfo{author}{\bibfnamefont{E.}~\bibnamefont{Mills}}, \bibinfo{author}{\bibfnamefont{L.}~\bibnamefont{Jones}}, \bibinfo{author}{\bibfnamefont{K.~R.} \bibnamefont{Symon}}, \bibnamefont{and} \bibinfo{author}{\bibfnamefont{D.}~\bibnamefont{Young}}, \emph{\bibinfo{title}{Innovation was not enough: a history of the {Midwestern} {Universities} {Research} {Association} ({MURA})}} (\bibinfo{publisher}{World Scientific}, \bibinfo{address}{Hackensack, N.J}, \bibinfo{year}{2010}), ISBN \bibinfo{isbn}{978-981-283-283-2}.

\bibitem[{\citenamefont{Ahdida et~al.}(2019)\citenamefont{Ahdida, Appleby, Bartmann, Bauche, Calviani, Gall, Gilardoni, Goddard, Hessler, Huber et~al.}}]{ahdida_nustorm_2019}
\bibinfo{author}{\bibfnamefont{C.}~\bibnamefont{Ahdida}}, \bibinfo{author}{\bibfnamefont{R.~B.} \bibnamefont{Appleby}}, \bibinfo{author}{\bibfnamefont{W.}~\bibnamefont{Bartmann}}, \bibinfo{author}{\bibfnamefont{J.}~\bibnamefont{Bauche}}, \bibinfo{author}{\bibfnamefont{M.}~\bibnamefont{Calviani}}, \bibinfo{author}{\bibfnamefont{J.}~\bibnamefont{Gall}}, \bibinfo{author}{\bibfnamefont{S.}~\bibnamefont{Gilardoni}}, \bibinfo{author}{\bibfnamefont{B.}~\bibnamefont{Goddard}}, \bibinfo{author}{\bibfnamefont{C.}~\bibnamefont{Hessler}}, \bibinfo{author}{\bibfnamefont{P.}~\bibnamefont{Huber}}, \bibnamefont{et~al.} (\bibinfo{year}{2019}), \bibinfo{note}{publisher: CERN Document Server}, \urlprefix\url{http://cds.cern.ch/record/2654649}.

\bibitem[{\citenamefont{Trbojevic et~al.}(2021{\natexlab{a}})\citenamefont{Trbojevic, Brooks, Hoffstaetter, Litvinenko, and Roser}}]{trbojevic_permanent_2021}
\bibinfo{author}{\bibfnamefont{D.}~\bibnamefont{Trbojevic}}, \bibinfo{author}{\bibfnamefont{S.}~\bibnamefont{Brooks}}, \bibinfo{author}{\bibfnamefont{G.}~\bibnamefont{Hoffstaetter}}, \bibinfo{author}{\bibfnamefont{V.}~\bibnamefont{Litvinenko}}, \bibnamefont{and} \bibinfo{author}{\bibfnamefont{T.}~\bibnamefont{Roser}}, \bibinfo{journal}{Proceedings of the 12th International Particle Accelerator Conference} \textbf{\bibinfo{volume}{IPAC2021}}, \bibinfo{pages}{4 pages, 1.741 MB} (\bibinfo{year}{2021}{\natexlab{a}}), ISSN \bibinfo{issn}{2673-5490}.

\bibitem[{\citenamefont{Peach et~al.}(2013)\citenamefont{Peach, Aslaninejad, Barlow, Beard, Bliss, Cobb, Easton, Edgecock, Fenning, Gardner et~al.}}]{peach_conceptual_2013}
\bibinfo{author}{\bibfnamefont{K.~J.} \bibnamefont{Peach}}, \bibinfo{author}{\bibfnamefont{M.}~\bibnamefont{Aslaninejad}}, \bibinfo{author}{\bibfnamefont{R.~J.} \bibnamefont{Barlow}}, \bibinfo{author}{\bibfnamefont{C.~D.} \bibnamefont{Beard}}, \bibinfo{author}{\bibfnamefont{N.}~\bibnamefont{Bliss}}, \bibinfo{author}{\bibfnamefont{J.~H.} \bibnamefont{Cobb}}, \bibinfo{author}{\bibfnamefont{M.~J.} \bibnamefont{Easton}}, \bibinfo{author}{\bibfnamefont{T.~R.} \bibnamefont{Edgecock}}, \bibinfo{author}{\bibfnamefont{R.}~\bibnamefont{Fenning}}, \bibinfo{author}{\bibfnamefont{I.~S.~K.} \bibnamefont{Gardner}}, \bibnamefont{et~al.}, \bibinfo{journal}{Physical Review Special Topics - Accelerators and Beams} \textbf{\bibinfo{volume}{16}}, \bibinfo{pages}{030101} (\bibinfo{year}{2013}), ISSN \bibinfo{issn}{1098-4402}, \urlprefix\url{https://link.aps.org/doi/10.1103/PhysRevSTAB.16.030101}.

\bibitem[{\citenamefont{Garland et~al.}(2015)\citenamefont{Garland, Appleby, Owen, and Tygier}}]{garland_normal-conducting_2015}
\bibinfo{author}{\bibfnamefont{J.}~\bibnamefont{Garland}}, \bibinfo{author}{\bibfnamefont{R.}~\bibnamefont{Appleby}}, \bibinfo{author}{\bibfnamefont{H.}~\bibnamefont{Owen}}, \bibnamefont{and} \bibinfo{author}{\bibfnamefont{S.}~\bibnamefont{Tygier}}, \bibinfo{journal}{Physical Review Special Topics - Accelerators and Beams} \textbf{\bibinfo{volume}{18}}, \bibinfo{pages}{094701} (\bibinfo{year}{2015}), ISSN \bibinfo{issn}{1098-4402}, \urlprefix\url{https://link.aps.org/doi/10.1103/PhysRevSTAB.18.094701}.

\bibitem[{\citenamefont{Meot}(2019)}]{meot_raccam_2019}
\bibinfo{author}{\bibfnamefont{F.}~\bibnamefont{Meot}}, \bibinfo{type}{Tech. Rep.} \bibinfo{number}{BNL--211536-2019-NEWS, 1507116}, \bibinfo{institution}{BNL} (\bibinfo{year}{2019}), \urlprefix\url{http://www.osti.gov/servlets/purl/1507116/}.

\bibitem[{\citenamefont{Craddock and Symon}(2008)}]{craddock_cyclotrons_2008}
\bibinfo{author}{\bibfnamefont{M.}~\bibnamefont{Craddock}} \bibnamefont{and} \bibinfo{author}{\bibfnamefont{K.}~\bibnamefont{Symon}}, \bibinfo{journal}{Reviews of Accelerator Science and Technology} \textbf{\bibinfo{volume}{01}} (\bibinfo{year}{2008}).

\bibitem[{\citenamefont{Méot}(1999)}]{meot_ray-tracing_1999}
\bibinfo{author}{\bibfnamefont{F.}~\bibnamefont{Méot}}, \bibinfo{journal}{Nuclear Instruments and Methods in Physics Research Section A: Accelerators, Spectrometers, Detectors and Associated Equipment} \textbf{\bibinfo{volume}{427}}, \bibinfo{pages}{353} (\bibinfo{year}{1999}), ISSN \bibinfo{issn}{01689002}, \urlprefix\url{https://linkinghub.elsevier.com/retrieve/pii/S0168900298015083}.

\bibitem[{\citenamefont{Vanwelde et~al.}(2023)\citenamefont{Vanwelde, Hernalsteens, Tesse, Gnacadja, Ramoisiaux, and Pauly}}]{vanwelde_zgoubidoo_2023}
\bibinfo{author}{\bibfnamefont{M.}~\bibnamefont{Vanwelde}}, \bibinfo{author}{\bibfnamefont{C.}~\bibnamefont{Hernalsteens}}, \bibinfo{author}{\bibfnamefont{R.}~\bibnamefont{Tesse}}, \bibinfo{author}{\bibfnamefont{E.}~\bibnamefont{Gnacadja}}, \bibinfo{author}{\bibfnamefont{E.}~\bibnamefont{Ramoisiaux}}, \bibnamefont{and} \bibinfo{author}{\bibfnamefont{N.}~\bibnamefont{Pauly}}, \bibinfo{journal}{Journal of Physics: Conference Series} \textbf{\bibinfo{volume}{2420}}, \bibinfo{pages}{012039} (\bibinfo{year}{2023}), ISSN \bibinfo{issn}{1742-6588, 1742-6596}, \urlprefix\url{https://iopscience.iop.org/article/10.1088/1742-6596/2420/1/012039}.

\bibitem[{\citenamefont{Tanigaki et~al.}(2006)\citenamefont{Tanigaki, Mori, Inoue, Mishima, Shiroya, Ishi, Fukumoto, and Machida}}]{tanigaki_present_2006}
\bibinfo{author}{\bibfnamefont{M.}~\bibnamefont{Tanigaki}}, \bibinfo{author}{\bibfnamefont{Y.}~\bibnamefont{Mori}}, \bibinfo{author}{\bibfnamefont{M.}~\bibnamefont{Inoue}}, \bibinfo{author}{\bibfnamefont{K.}~\bibnamefont{Mishima}}, \bibinfo{author}{\bibfnamefont{S.}~\bibnamefont{Shiroya}}, \bibinfo{author}{\bibfnamefont{Y.}~\bibnamefont{Ishi}}, \bibinfo{author}{\bibfnamefont{S.}~\bibnamefont{Fukumoto}}, \bibnamefont{and} \bibinfo{author}{\bibfnamefont{S.}~\bibnamefont{Machida}}, \bibinfo{journal}{Proceedings of EPAC 2006}  (\bibinfo{year}{2006}).

\bibitem[{\citenamefont{Suga et~al.}(2019)\citenamefont{Suga, Ishi, Uesugi, Kuriyama, Fuwa, Okita, and Mori}}]{suga_remodeling_2019}
\bibinfo{author}{\bibfnamefont{K.}~\bibnamefont{Suga}}, \bibinfo{author}{\bibfnamefont{Y.}~\bibnamefont{Ishi}}, \bibinfo{author}{\bibfnamefont{T.}~\bibnamefont{Uesugi}}, \bibinfo{author}{\bibfnamefont{Y.}~\bibnamefont{Kuriyama}}, \bibinfo{author}{\bibfnamefont{Y.}~\bibnamefont{Fuwa}}, \bibinfo{author}{\bibfnamefont{H.}~\bibnamefont{Okita}}, \bibnamefont{and} \bibinfo{author}{\bibfnamefont{Y.}~\bibnamefont{Mori}}, \bibinfo{journal}{Journal of Physics: Conference Series} \textbf{\bibinfo{volume}{1350}}, \bibinfo{pages}{012070} (\bibinfo{year}{2019}), ISSN \bibinfo{issn}{1742-6588, 1742-6596}, \urlprefix\url{https://iopscience.iop.org/article/10.1088/1742-6596/1350/1/012070}.

\bibitem[{\citenamefont{Okita et~al.}(2019)\citenamefont{Okita, Taniguchi, Kuriyama, Uesugi, Ishi, Mori, Muto, Ono, Ikeda, Yonemura et~al.}}]{okita_beam_2019}
\bibinfo{author}{\bibfnamefont{H.}~\bibnamefont{Okita}}, \bibinfo{author}{\bibfnamefont{A.}~\bibnamefont{Taniguchi}}, \bibinfo{author}{\bibfnamefont{Y.}~\bibnamefont{Kuriyama}}, \bibinfo{author}{\bibfnamefont{T.}~\bibnamefont{Uesugi}}, \bibinfo{author}{\bibfnamefont{Y.}~\bibnamefont{Ishi}}, \bibinfo{author}{\bibfnamefont{Y.}~\bibnamefont{Mori}}, \bibinfo{author}{\bibfnamefont{M.}~\bibnamefont{Muto}}, \bibinfo{author}{\bibfnamefont{Y.}~\bibnamefont{Ono}}, \bibinfo{author}{\bibfnamefont{N.}~\bibnamefont{Ikeda}}, \bibinfo{author}{\bibfnamefont{Y.}~\bibnamefont{Yonemura}}, \bibnamefont{et~al.}, \bibinfo{journal}{Journal of Physics: Conference Series} \textbf{\bibinfo{volume}{1350}}, \bibinfo{pages}{012069} (\bibinfo{year}{2019}), ISSN \bibinfo{issn}{1742-6588, 1742-6596}, \urlprefix\url{https://iopscience.iop.org/article/10.1088/1742-6596/1350/1/012069}.

\bibitem[{\citenamefont{Machida et~al.}(2012)\citenamefont{Machida, Barlow, Berg, Bliss, Buckley, Clarke, Craddock, D'Arcy, Edgecock, Garland et~al.}}]{machida_acceleration_2012}
\bibinfo{author}{\bibfnamefont{S.}~\bibnamefont{Machida}}, \bibinfo{author}{\bibfnamefont{R.}~\bibnamefont{Barlow}}, \bibinfo{author}{\bibfnamefont{J.~S.} \bibnamefont{Berg}}, \bibinfo{author}{\bibfnamefont{N.}~\bibnamefont{Bliss}}, \bibinfo{author}{\bibfnamefont{R.~K.} \bibnamefont{Buckley}}, \bibinfo{author}{\bibfnamefont{J.~A.} \bibnamefont{Clarke}}, \bibinfo{author}{\bibfnamefont{M.~K.} \bibnamefont{Craddock}}, \bibinfo{author}{\bibfnamefont{R.}~\bibnamefont{D'Arcy}}, \bibinfo{author}{\bibfnamefont{R.}~\bibnamefont{Edgecock}}, \bibinfo{author}{\bibfnamefont{J.~M.} \bibnamefont{Garland}}, \bibnamefont{et~al.}, \bibinfo{journal}{Nature Physics} \textbf{\bibinfo{volume}{8}}, \bibinfo{pages}{243} (\bibinfo{year}{2012}), ISSN \bibinfo{issn}{1745-2473, 1745-2481}, \urlprefix\url{http://www.nature.com/articles/nphys2179}.

\bibitem[{\citenamefont{Machida et~al.}(2023)\citenamefont{Machida, Letchford, Jolly, Rogers, Kelliher, Yamakawa, Pasternak, and Lagrange}}]{machida_ffa_2023}
\bibinfo{author}{\bibfnamefont{S.}~\bibnamefont{Machida}}, \bibinfo{author}{\bibfnamefont{A.}~\bibnamefont{Letchford}}, \bibinfo{author}{\bibfnamefont{C.}~\bibnamefont{Jolly}}, \bibinfo{author}{\bibfnamefont{C.}~\bibnamefont{Rogers}}, \bibinfo{author}{\bibfnamefont{D.}~\bibnamefont{Kelliher}}, \bibinfo{author}{\bibfnamefont{E.}~\bibnamefont{Yamakawa}}, \bibinfo{author}{\bibfnamefont{J.}~\bibnamefont{Pasternak}}, \bibnamefont{and} \bibinfo{author}{\bibfnamefont{J.}~\bibnamefont{Lagrange}}, \bibinfo{journal}{Proceedings of the 14th International Particle Accelerator Conference}  (\bibinfo{year}{2023}).

\bibitem[{\citenamefont{Benesch et~al.}(2023)\citenamefont{Benesch, Berg, Bodenstein, Bogacz, Brooks, Coxe, Deitrick, Gamage, Hoffstaetter, Khan et~al.}}]{benesch_cebaf_2023}
\bibinfo{author}{\bibfnamefont{J.}~\bibnamefont{Benesch}}, \bibinfo{author}{\bibfnamefont{J.}~\bibnamefont{Berg}}, \bibinfo{author}{\bibfnamefont{R.}~\bibnamefont{Bodenstein}}, \bibinfo{author}{\bibfnamefont{A.}~\bibnamefont{Bogacz}}, \bibinfo{author}{\bibfnamefont{S.}~\bibnamefont{Brooks}}, \bibinfo{author}{\bibfnamefont{A.}~\bibnamefont{Coxe}}, \bibinfo{author}{\bibfnamefont{K.}~\bibnamefont{Deitrick}}, \bibinfo{author}{\bibfnamefont{B.}~\bibnamefont{Gamage}}, \bibinfo{author}{\bibfnamefont{G.}~\bibnamefont{Hoffstaetter}}, \bibinfo{author}{\bibfnamefont{D.}~\bibnamefont{Khan}}, \bibnamefont{et~al.}, \bibinfo{journal}{Proceedings of the 14th International Particle Accelerator Conference}  (\bibinfo{year}{2023}).

\bibitem[{\citenamefont{Johnstone}(2023)}]{johnstone_new_2023}
\bibinfo{author}{\bibfnamefont{C.}~\bibnamefont{Johnstone}}, \bibinfo{journal}{Proceedings of the 14th International Particle Accelerator Conference}  (\bibinfo{year}{2023}).

\bibitem[{\citenamefont{Aymar et~al.}(2020)\citenamefont{Aymar, Becker, Boogert, Borghesi, Bingham, Brenner, Burrows, Ettlinger, Dascalu, Gibson et~al.}}]{aymar_lhara_2020}
\bibinfo{author}{\bibfnamefont{G.}~\bibnamefont{Aymar}}, \bibinfo{author}{\bibfnamefont{T.}~\bibnamefont{Becker}}, \bibinfo{author}{\bibfnamefont{S.}~\bibnamefont{Boogert}}, \bibinfo{author}{\bibfnamefont{M.}~\bibnamefont{Borghesi}}, \bibinfo{author}{\bibfnamefont{R.}~\bibnamefont{Bingham}}, \bibinfo{author}{\bibfnamefont{C.}~\bibnamefont{Brenner}}, \bibinfo{author}{\bibfnamefont{P.~N.} \bibnamefont{Burrows}}, \bibinfo{author}{\bibfnamefont{O.~C.} \bibnamefont{Ettlinger}}, \bibinfo{author}{\bibfnamefont{T.}~\bibnamefont{Dascalu}}, \bibinfo{author}{\bibfnamefont{S.}~\bibnamefont{Gibson}}, \bibnamefont{et~al.}, \bibinfo{journal}{Frontiers in Physics} \textbf{\bibinfo{volume}{8}}, \bibinfo{pages}{567738} (\bibinfo{year}{2020}), ISSN \bibinfo{issn}{2296-424X}, \urlprefix\url{https://www.frontiersin.org/articles/10.3389/fphy.2020.567738/full}.

\bibitem[{\citenamefont{Wiedemann}(2015)}]{wiedemann_particle_2015}
\bibinfo{author}{\bibfnamefont{H.}~\bibnamefont{Wiedemann}}, \emph{\bibinfo{title}{Particle {Accelerator} {Physics}}}, Graduate {Texts} in {Physics} (\bibinfo{publisher}{Springer International Publishing}, \bibinfo{address}{Cham}, \bibinfo{year}{2015}), ISBN \bibinfo{isbn}{978-3-319-18316-9 978-3-319-18317-6}, \urlprefix\url{http://link.springer.com/10.1007/978-3-319-18317-6}.

\bibitem[{\citenamefont{Holzer}(2014)}]{holzer_lattice_2014}
\bibinfo{author}{\bibfnamefont{B.}~\bibnamefont{Holzer}} (\bibinfo{year}{2014}), \bibinfo{note}{publisher: CERN}, \urlprefix\url{https://cds.cern.ch/record/1982419}.

\bibitem[{\citenamefont{Bodenstein et~al.}(2023)\citenamefont{Bodenstein, Bogacz, Coxe, Seryi, Gamage, Trbojevic, Khan, Berg, Benesch, Price et~al.}}]{bodenstein_designing_2023}
\bibinfo{author}{\bibfnamefont{R.}~\bibnamefont{Bodenstein}}, \bibinfo{author}{\bibfnamefont{A.}~\bibnamefont{Bogacz}}, \bibinfo{author}{\bibfnamefont{A.}~\bibnamefont{Coxe}}, \bibinfo{author}{\bibfnamefont{A.}~\bibnamefont{Seryi}}, \bibinfo{author}{\bibfnamefont{B.}~\bibnamefont{Gamage}}, \bibinfo{author}{\bibfnamefont{D.}~\bibnamefont{Trbojevic}}, \bibinfo{author}{\bibfnamefont{D.}~\bibnamefont{Khan}}, \bibinfo{author}{\bibfnamefont{J.}~\bibnamefont{Berg}}, \bibinfo{author}{\bibfnamefont{J.}~\bibnamefont{Benesch}}, \bibinfo{author}{\bibfnamefont{K.}~\bibnamefont{Price}}, \bibnamefont{et~al.}, \bibinfo{journal}{Proceedings of the 14th International Particle Accelerator Conference}  (\bibinfo{year}{2023}).

\bibitem[{\citenamefont{Dascalu and Sheehy}(2021)}]{dascalu_beam_2021}
\bibinfo{author}{\bibfnamefont{T.-S.} \bibnamefont{Dascalu}} \bibnamefont{and} \bibinfo{author}{\bibfnamefont{S.~L.} \bibnamefont{Sheehy}}, \emph{\bibinfo{title}{Beam delivery systems for linac-based proton therapy}} (\bibinfo{year}{2021}), \bibinfo{note}{arXiv:2103.05315 [physics]}, \urlprefix\url{http://arxiv.org/abs/2103.05315}.

\bibitem[{\citenamefont{Zhao et~al.}(2020)\citenamefont{Zhao, Qin, Liu, Chen, and Chen}}]{zhao_design_2020}
\bibinfo{author}{\bibfnamefont{R.}~\bibnamefont{Zhao}}, \bibinfo{author}{\bibfnamefont{B.}~\bibnamefont{Qin}}, \bibinfo{author}{\bibfnamefont{X.}~\bibnamefont{Liu}}, \bibinfo{author}{\bibfnamefont{H.}~\bibnamefont{Chen}}, \bibnamefont{and} \bibinfo{author}{\bibfnamefont{Q.}~\bibnamefont{Chen}}, \bibinfo{journal}{Physica Medica} \textbf{\bibinfo{volume}{73}}, \bibinfo{pages}{158} (\bibinfo{year}{2020}), ISSN \bibinfo{issn}{11201797}, \urlprefix\url{https://linkinghub.elsevier.com/retrieve/pii/S1120179720301010}.

\bibitem[{\citenamefont{Nesteruk et~al.}(2019)\citenamefont{Nesteruk, Calzolaio, Meer, Rizzoglio, Seidel, and Schippers}}]{nesteruk_large_2019}
\bibinfo{author}{\bibfnamefont{K.~P.} \bibnamefont{Nesteruk}}, \bibinfo{author}{\bibfnamefont{C.}~\bibnamefont{Calzolaio}}, \bibinfo{author}{\bibfnamefont{D.}~\bibnamefont{Meer}}, \bibinfo{author}{\bibfnamefont{V.}~\bibnamefont{Rizzoglio}}, \bibinfo{author}{\bibfnamefont{M.}~\bibnamefont{Seidel}}, \bibnamefont{and} \bibinfo{author}{\bibfnamefont{J.~M.} \bibnamefont{Schippers}}, \bibinfo{journal}{Physics in Medicine \& Biology} \textbf{\bibinfo{volume}{64}}, \bibinfo{pages}{175007} (\bibinfo{year}{2019}), ISSN \bibinfo{issn}{1361-6560}, \urlprefix\url{https://iopscience.iop.org/article/10.1088/1361-6560/ab2f5f}.

\bibitem[{\citenamefont{Pasternak et~al.}(2013)\citenamefont{Pasternak, Aslaninejad, Holland, Posocco, and Walton}}]{pasternak_novel_2013}
\bibinfo{author}{\bibfnamefont{J.}~\bibnamefont{Pasternak}}, \bibinfo{author}{\bibfnamefont{M.}~\bibnamefont{Aslaninejad}}, \bibinfo{author}{\bibfnamefont{P.~R.~N.} \bibnamefont{Holland}}, \bibinfo{author}{\bibfnamefont{P.~A.} \bibnamefont{Posocco}}, \bibnamefont{and} \bibinfo{author}{\bibfnamefont{G.~W.} \bibnamefont{Walton}}, \bibinfo{journal}{Proceedings of PAC2013} p.~\bibinfo{pages}{3} (\bibinfo{year}{2013}).

\bibitem[{\citenamefont{Brouwer et~al.}(2019)\citenamefont{Brouwer, Huggins, and Wan}}]{brouwer_achromatic_2019}
\bibinfo{author}{\bibfnamefont{L.}~\bibnamefont{Brouwer}}, \bibinfo{author}{\bibfnamefont{A.}~\bibnamefont{Huggins}}, \bibnamefont{and} \bibinfo{author}{\bibfnamefont{W.}~\bibnamefont{Wan}}, \bibinfo{journal}{International Journal of Modern Physics A} \textbf{\bibinfo{volume}{34}}, \bibinfo{pages}{1942023} (\bibinfo{year}{2019}), ISSN \bibinfo{issn}{0217-751X, 1793-656X}, \urlprefix\url{https://www.worldscientific.com/doi/abs/10.1142/S0217751X19420235}.

\bibitem[{\citenamefont{Trbojevic et~al.}(2021{\natexlab{b}})\citenamefont{Trbojevic, Brooks, Roser, and Tsoupas}}]{trbojevic_superb_2021}
\bibinfo{author}{\bibfnamefont{D.}~\bibnamefont{Trbojevic}}, \bibinfo{author}{\bibfnamefont{S.}~\bibnamefont{Brooks}}, \bibinfo{author}{\bibfnamefont{T.}~\bibnamefont{Roser}}, \bibnamefont{and} \bibinfo{author}{\bibfnamefont{N.}~\bibnamefont{Tsoupas}}, \bibinfo{journal}{Proceedings of the 12th International Particle Accelerator Conference} \textbf{\bibinfo{volume}{IPAC2021}}, \bibinfo{pages}{4 pages, 1.677 MB} (\bibinfo{year}{2021}{\natexlab{b}}), ISSN \bibinfo{issn}{2673-5490}, \urlprefix\url{https://jacow.org/ipac2021/doi/JACoW-IPAC2021-TUPAB030.html}.

\bibitem[{\citenamefont{Tesse et~al.}(2023{\natexlab{a}})\citenamefont{Tesse, Hernalsteens, Ramoisiaux, Gnacadja, Vanwelde, and Pauly}}]{tesse_gantry_2023}
\bibinfo{author}{\bibfnamefont{R.}~\bibnamefont{Tesse}}, \bibinfo{author}{\bibfnamefont{C.}~\bibnamefont{Hernalsteens}}, \bibinfo{author}{\bibfnamefont{E.}~\bibnamefont{Ramoisiaux}}, \bibinfo{author}{\bibfnamefont{E.}~\bibnamefont{Gnacadja}}, \bibinfo{author}{\bibfnamefont{M.}~\bibnamefont{Vanwelde}}, \bibnamefont{and} \bibinfo{author}{\bibfnamefont{N.}~\bibnamefont{Pauly}}, \bibinfo{journal}{Proceedings of the 14th International Particle Accelerator Conference}  (\bibinfo{year}{2023}{\natexlab{a}}).

\bibitem[{\citenamefont{Tesse et~al.}(2023{\natexlab{b}})\citenamefont{Tesse, Hernalsteens, Gnacadja, Ramoisiaux, Pauly, and Vanwelde}}]{tesse_achromatic_2023}
\bibinfo{author}{\bibfnamefont{R.}~\bibnamefont{Tesse}}, \bibinfo{author}{\bibfnamefont{C.}~\bibnamefont{Hernalsteens}}, \bibinfo{author}{\bibfnamefont{E.}~\bibnamefont{Gnacadja}}, \bibinfo{author}{\bibfnamefont{E.}~\bibnamefont{Ramoisiaux}}, \bibinfo{author}{\bibfnamefont{N.}~\bibnamefont{Pauly}}, \bibnamefont{and} \bibinfo{author}{\bibfnamefont{M.}~\bibnamefont{Vanwelde}}, \bibinfo{journal}{Journal of Physics: Conference Series} \textbf{\bibinfo{volume}{2420}}, \bibinfo{pages}{012096} (\bibinfo{year}{2023}{\natexlab{b}}), ISSN \bibinfo{issn}{1742-6588, 1742-6596}, \urlprefix\url{https://iopscience.iop.org/article/10.1088/1742-6596/2420/1/012096}.

\bibitem[{\citenamefont{Machida and Fenning}(2010)}]{machida_beam_2010}
\bibinfo{author}{\bibfnamefont{S.}~\bibnamefont{Machida}} \bibnamefont{and} \bibinfo{author}{\bibfnamefont{R.}~\bibnamefont{Fenning}}, \bibinfo{journal}{Physical Review Special Topics - Accelerators and Beams} \textbf{\bibinfo{volume}{13}}, \bibinfo{pages}{084001} (\bibinfo{year}{2010}), ISSN \bibinfo{issn}{1098-4402}, \urlprefix\url{https://link.aps.org/doi/10.1103/PhysRevSTAB.13.084001}.

\bibitem[{\citenamefont{Fenning et~al.}(2012)\citenamefont{Fenning, Machida, Kelliher, Khan, and Edgecock}}]{fenning_high-order_2012}
\bibinfo{author}{\bibfnamefont{R.}~\bibnamefont{Fenning}}, \bibinfo{author}{\bibfnamefont{S.}~\bibnamefont{Machida}}, \bibinfo{author}{\bibfnamefont{D.}~\bibnamefont{Kelliher}}, \bibinfo{author}{\bibfnamefont{A.}~\bibnamefont{Khan}}, \bibnamefont{and} \bibinfo{author}{\bibfnamefont{R.}~\bibnamefont{Edgecock}}, \bibinfo{journal}{Journal of Instrumentation} \textbf{\bibinfo{volume}{7}}, \bibinfo{pages}{P05011} (\bibinfo{year}{2012}), ISSN \bibinfo{issn}{1748-0221}, \urlprefix\url{https://iopscience.iop.org/article/10.1088/1748-0221/7/05/P05011}.

\bibitem[{\citenamefont{Shepherd}(2020)}]{shepherd_permanent_2020}
\bibinfo{author}{\bibfnamefont{B.}~\bibnamefont{Shepherd}}, \bibinfo{journal}{Proceedings of the 11th International Particle Accelerator Conference} \textbf{\bibinfo{volume}{IPAC2020}}, \bibinfo{pages}{5 pages, 1.467 MB} (\bibinfo{year}{2020}), ISSN \bibinfo{issn}{2673-5490}, \urlprefix\url{https://jacow.org/ipac2020/doi/JACoW-IPAC2020-MOVIRO05.html}.

\bibitem[{\citenamefont{Halbach}(1980)}]{halbach_design_1980}
\bibinfo{author}{\bibfnamefont{K.}~\bibnamefont{Halbach}}, \bibinfo{journal}{Nuclear Instruments and Methods} \textbf{\bibinfo{volume}{169}}, \bibinfo{pages}{1} (\bibinfo{year}{1980}), ISSN \bibinfo{issn}{0029554X}, \urlprefix\url{https://linkinghub.elsevier.com/retrieve/pii/0029554X80900944}.

\bibitem[{\citenamefont{Clarke}(2004)}]{clarke_science_2004}
\bibinfo{author}{\bibfnamefont{J.~A.} \bibnamefont{Clarke}}, \emph{\bibinfo{title}{The {Science} and {Technology} of {Undulators} and {Wigglers}}} (\bibinfo{publisher}{Oxford University Press}, \bibinfo{year}{2004}), ISBN \bibinfo{isbn}{978-0-19-850855-7}, \urlprefix\url{https://academic.oup.com/book/4976}.

\bibitem[{\citenamefont{Thonet}(2016)}]{thonet_use_2016}
\bibinfo{author}{\bibfnamefont{P.}~\bibnamefont{Thonet}}, \bibinfo{journal}{IEEE Transactions on Applied Superconductivity} pp. \bibinfo{pages}{1--1} (\bibinfo{year}{2016}), ISSN \bibinfo{issn}{1051-8223, 1558-2515}, \urlprefix\url{http://ieeexplore.ieee.org/document/7414437/}.

\bibitem[{\citenamefont{Wangler}(2008)}]{wangler_rf_2008}
\bibinfo{author}{\bibfnamefont{T.~P.} \bibnamefont{Wangler}}, \emph{\bibinfo{title}{{RF} {Linear} {Accelerators}}} (\bibinfo{publisher}{Wiley}, \bibinfo{year}{2008}), \bibinfo{edition}{1st} ed., ISBN \bibinfo{isbn}{978-3-527-40680-7 978-3-527-62342-6}, \urlprefix\url{https://onlinelibrary.wiley.com/doi/book/10.1002/9783527623426}.

\bibitem[{\citenamefont{Shepherd et~al.}(2014)\citenamefont{Shepherd, Clarke, Marks, Collomb, Stokes, Modena, Struik, and Bartalesi}}]{shepherd_tunable_2014}
\bibinfo{author}{\bibfnamefont{B.}~\bibnamefont{Shepherd}}, \bibinfo{author}{\bibfnamefont{J.}~\bibnamefont{Clarke}}, \bibinfo{author}{\bibfnamefont{N.}~\bibnamefont{Marks}}, \bibinfo{author}{\bibfnamefont{N.}~\bibnamefont{Collomb}}, \bibinfo{author}{\bibfnamefont{D.}~\bibnamefont{Stokes}}, \bibinfo{author}{\bibfnamefont{M.}~\bibnamefont{Modena}}, \bibinfo{author}{\bibfnamefont{M.}~\bibnamefont{Struik}}, \bibnamefont{and} \bibinfo{author}{\bibfnamefont{A.}~\bibnamefont{Bartalesi}}, \bibinfo{journal}{Journal of Instrumentation} \textbf{\bibinfo{volume}{9}}, \bibinfo{pages}{T11006} (\bibinfo{year}{2014}), ISSN \bibinfo{issn}{1748-0221}, \urlprefix\url{https://iopscience.iop.org/article/10.1088/1748-0221/9/11/T11006}.

\bibitem[{\citenamefont{Brooks et~al.}(2019)\citenamefont{Brooks, Mahler, Tuozzolo, and Michnoff}}]{brooks_cbeta_2019}
\bibinfo{author}{\bibfnamefont{S.}~\bibnamefont{Brooks}}, \bibinfo{author}{\bibfnamefont{G.}~\bibnamefont{Mahler}}, \bibinfo{author}{\bibfnamefont{J.}~\bibnamefont{Tuozzolo}}, \bibnamefont{and} \bibinfo{author}{\bibfnamefont{R.}~\bibnamefont{Michnoff}}, \bibinfo{journal}{10th International Particle Accelerator Conference (IPAC19)}  (\bibinfo{year}{2019}).

\bibitem[{\citenamefont{Brooks and Bogacz}(2022)}]{brooks_permanent_2022}
\bibinfo{author}{\bibfnamefont{S.}~\bibnamefont{Brooks}} \bibnamefont{and} \bibinfo{author}{\bibfnamefont{A.}~\bibnamefont{Bogacz}}, \bibinfo{journal}{Proceedings of the 13th International Particle Accelerator Conference} \textbf{\bibinfo{volume}{IPAC2022}}, \bibinfo{pages}{4 pages, 0.715 MB} (\bibinfo{year}{2022}), ISSN \bibinfo{issn}{2673-5490}, \urlprefix\url{https://jacow.org/ipac2022/doi/JACoW-IPAC2022-THPOTK011.html}.

\bibitem[{\citenamefont{Brooks et~al.}(2020)\citenamefont{Brooks, Mahler, Cintorino, Tuozzolo, and Michnoff}}]{brooks_permanent_2020}
\bibinfo{author}{\bibfnamefont{S.}~\bibnamefont{Brooks}}, \bibinfo{author}{\bibfnamefont{G.}~\bibnamefont{Mahler}}, \bibinfo{author}{\bibfnamefont{J.}~\bibnamefont{Cintorino}}, \bibinfo{author}{\bibfnamefont{J.}~\bibnamefont{Tuozzolo}}, \bibnamefont{and} \bibinfo{author}{\bibfnamefont{R.}~\bibnamefont{Michnoff}}, \bibinfo{journal}{Physical Review Accelerators and Beams} \textbf{\bibinfo{volume}{23}}, \bibinfo{pages}{112401} (\bibinfo{year}{2020}), ISSN \bibinfo{issn}{2469-9888}, \urlprefix\url{https://link.aps.org/doi/10.1103/PhysRevAccelBeams.23.112401}.

\bibitem[{noa(2024)}]{noauthor_100mm_2024}
\emph{\bibinfo{title}{100mm x 12.7mm x 12.7mm - {Fastmag} {Magnets}}} (\bibinfo{year}{2024}), \urlprefix\url{https://www.fastmag.com.au/product/re10012-712-7- rare-earth-block- 100mm-x-12-7mm-x-12-7mm}.

\bibitem[{\citenamefont{Ortner and Coliado Bandeira}(2020)}]{ortner_magpylib_2020}
\bibinfo{author}{\bibfnamefont{M.}~\bibnamefont{Ortner}} \bibnamefont{and} \bibinfo{author}{\bibfnamefont{L.~G.} \bibnamefont{Coliado Bandeira}}, \bibinfo{journal}{SoftwareX} \textbf{\bibinfo{volume}{11}}, \bibinfo{pages}{100466} (\bibinfo{year}{2020}), ISSN \bibinfo{issn}{23527110}, \urlprefix\url{https://linkinghub.elsevier.com/retrieve/pii/S2352711020300170}.

\bibitem[{\citenamefont{Bottura et~al.}(2016)\citenamefont{Bottura, Gourlay, Yamamoto, and Zlobin}}]{bottura_superconducting_2016}
\bibinfo{author}{\bibfnamefont{L.}~\bibnamefont{Bottura}}, \bibinfo{author}{\bibfnamefont{S.~A.} \bibnamefont{Gourlay}}, \bibinfo{author}{\bibfnamefont{A.}~\bibnamefont{Yamamoto}}, \bibnamefont{and} \bibinfo{author}{\bibfnamefont{A.~V.} \bibnamefont{Zlobin}}, \bibinfo{journal}{IEEE Transactions on Nuclear Science} \textbf{\bibinfo{volume}{63}}, \bibinfo{pages}{751} (\bibinfo{year}{2016}), ISSN \bibinfo{issn}{1558-1578}.

\bibitem[{\citenamefont{Jeans}(1911)}]{jeans_mathematical_1911}
\bibinfo{author}{\bibfnamefont{J.}~\bibnamefont{Jeans}}, \emph{\bibinfo{title}{Mathematical {Theory} of {Electricity} and {Magnetism}}} (\bibinfo{publisher}{Cambridge University Press}, \bibinfo{year}{1911}), \bibinfo{edition}{2nd} ed., ISBN \bibinfo{isbn}{978-1-108-00561-6 978-0-511-69435-6}, \urlprefix\url{https://www.cambridge.org/core/product/identifier/9780511694356/type/book}.

\bibitem[{\citenamefont{Prat-Camps et~al.}(2018)\citenamefont{Prat-Camps, Maurer, Kirchmair, and Romero-Isart}}]{prat-camps_circumventing_2018}
\bibinfo{author}{\bibfnamefont{J.}~\bibnamefont{Prat-Camps}}, \bibinfo{author}{\bibfnamefont{P.}~\bibnamefont{Maurer}}, \bibinfo{author}{\bibfnamefont{G.}~\bibnamefont{Kirchmair}}, \bibnamefont{and} \bibinfo{author}{\bibfnamefont{O.}~\bibnamefont{Romero-Isart}}, \bibinfo{journal}{Physical Review Letters} \textbf{\bibinfo{volume}{121}}, \bibinfo{pages}{213903} (\bibinfo{year}{2018}), ISSN \bibinfo{issn}{0031-9007, 1079-7114}, \urlprefix\url{https://link.aps.org/doi/10.1103/PhysRevLett.121.213903}.

\bibitem[{\citenamefont{Russenschuck}(2022)}]{russenschuck_design_2022}
\bibinfo{author}{\bibfnamefont{S.}~\bibnamefont{Russenschuck}} (\bibinfo{year}{2022}), \urlprefix\url{http://cds.cern.ch/record/865932}.

\bibitem[{\citenamefont{Enge}(1967)}]{enge_deflecting_1967}
\bibinfo{author}{\bibfnamefont{H.~A.} \bibnamefont{Enge}}, in \emph{\bibinfo{booktitle}{Focusing of {Charged} {Particles}}}, edited by \bibinfo{editor}{\bibfnamefont{S.}~\bibnamefont{Albert}} (\bibinfo{publisher}{Elsevier}, \bibinfo{year}{1967}), vol.~\bibinfo{volume}{2}, pp. \bibinfo{pages}{203--264}, ISBN \bibinfo{isbn}{978-0-12-636902-1}, \urlprefix\url{https://linkinghub.elsevier.com/retrieve/pii/B9780126369021500123}.

\bibitem[{\citenamefont{Sheehy et~al.}(2010)\citenamefont{Sheehy, Peach, Witte, Kelliher, and Machida}}]{sheehy_fixed_2010}
\bibinfo{author}{\bibfnamefont{S.~L.} \bibnamefont{Sheehy}}, \bibinfo{author}{\bibfnamefont{K.~J.} \bibnamefont{Peach}}, \bibinfo{author}{\bibfnamefont{H.}~\bibnamefont{Witte}}, \bibinfo{author}{\bibfnamefont{D.~J.} \bibnamefont{Kelliher}}, \bibnamefont{and} \bibinfo{author}{\bibfnamefont{S.}~\bibnamefont{Machida}}, \bibinfo{journal}{Physical Review Special Topics - Accelerators and Beams} \textbf{\bibinfo{volume}{13}}, \bibinfo{pages}{040101} (\bibinfo{year}{2010}), ISSN \bibinfo{issn}{1098-4402}, \urlprefix\url{https://link.aps.org/doi/10.1103/PhysRevSTAB.13.040101}.

\bibitem[{\citenamefont{Shepherd}(2018)}]{shepherd_radiation_2018}
\bibinfo{author}{\bibfnamefont{B.}~\bibnamefont{Shepherd}}, \bibinfo{journal}{CLIC Technical Note} \textbf{\bibinfo{volume}{1079}} (\bibinfo{year}{2018}).

\bibitem[{\citenamefont{Machida}(2009)}]{machida_scaling_2009}
\bibinfo{author}{\bibfnamefont{S.}~\bibnamefont{Machida}}, \bibinfo{journal}{Physical Review Letters} \textbf{\bibinfo{volume}{103}}, \bibinfo{pages}{164801} (\bibinfo{year}{2009}), ISSN \bibinfo{issn}{0031-9007, 1079-7114}, \urlprefix\url{https://link.aps.org/doi/10.1103/PhysRevLett.103.164801}.

\bibitem[{\citenamefont{Blank and Deb}(2020)}]{blank_pymoo_2020}
\bibinfo{author}{\bibfnamefont{J.}~\bibnamefont{Blank}} \bibnamefont{and} \bibinfo{author}{\bibfnamefont{K.}~\bibnamefont{Deb}}, \bibinfo{journal}{IEEE Access} \textbf{\bibinfo{volume}{8}}, \bibinfo{pages}{89497} (\bibinfo{year}{2020}), ISSN \bibinfo{issn}{2169-3536}, \bibinfo{note}{conference Name: IEEE Access}.

\bibitem[{\citenamefont{Zhao et~al.}(2021)\citenamefont{Zhao, Qin, Liao, Liu, Chen, and Han}}]{zhao_achieving_2021}
\bibinfo{author}{\bibfnamefont{R.}~\bibnamefont{Zhao}}, \bibinfo{author}{\bibfnamefont{B.}~\bibnamefont{Qin}}, \bibinfo{author}{\bibfnamefont{Y.}~\bibnamefont{Liao}}, \bibinfo{author}{\bibfnamefont{X.}~\bibnamefont{Liu}}, \bibinfo{author}{\bibfnamefont{Q.}~\bibnamefont{Chen}}, \bibnamefont{and} \bibinfo{author}{\bibfnamefont{W.}~\bibnamefont{Han}}, \bibinfo{journal}{Nuclear Instruments and Methods in Physics Research Section A: Accelerators, Spectrometers, Detectors and Associated Equipment} \textbf{\bibinfo{volume}{1015}}, \bibinfo{pages}{165773} (\bibinfo{year}{2021}), ISSN \bibinfo{issn}{01689002}, \urlprefix\url{https://linkinghub.elsevier.com/retrieve/pii/S0168900221007580}.

\bibitem[{\citenamefont{Arjomandy et~al.}(2019)\citenamefont{Arjomandy, Taylor, Ainsley, Safai, Sahoo, Pankuch, Farr, Yong~Park, Klein, Flanz et~al.}}]{arjomandy_aapm_2019}
\bibinfo{author}{\bibfnamefont{B.}~\bibnamefont{Arjomandy}}, \bibinfo{author}{\bibfnamefont{P.}~\bibnamefont{Taylor}}, \bibinfo{author}{\bibfnamefont{C.}~\bibnamefont{Ainsley}}, \bibinfo{author}{\bibfnamefont{S.}~\bibnamefont{Safai}}, \bibinfo{author}{\bibfnamefont{N.}~\bibnamefont{Sahoo}}, \bibinfo{author}{\bibfnamefont{M.}~\bibnamefont{Pankuch}}, \bibinfo{author}{\bibfnamefont{J.~B.} \bibnamefont{Farr}}, \bibinfo{author}{\bibfnamefont{S.}~\bibnamefont{Yong~Park}}, \bibinfo{author}{\bibfnamefont{E.}~\bibnamefont{Klein}}, \bibinfo{author}{\bibfnamefont{J.}~\bibnamefont{Flanz}}, \bibnamefont{et~al.}, \bibinfo{journal}{Medical Physics} \textbf{\bibinfo{volume}{46}}, \bibinfo{pages}{e678} (\bibinfo{year}{2019}), ISSN \bibinfo{issn}{2473-4209}, \bibinfo{note}{\_eprint: https://onlinelibrary.wiley.com/doi/pdf/10.1002/mp.13622}, \urlprefix\url{https://onlinelibrary.wiley.com/doi/abs/10.1002/mp.13622}.

\bibitem[{\citenamefont{Safai et~al.}(2008)\citenamefont{Safai, Bortfeld, and Engelsman}}]{safai_comparison_2008}
\bibinfo{author}{\bibfnamefont{S.}~\bibnamefont{Safai}}, \bibinfo{author}{\bibfnamefont{T.}~\bibnamefont{Bortfeld}}, \bibnamefont{and} \bibinfo{author}{\bibfnamefont{M.}~\bibnamefont{Engelsman}}, \bibinfo{journal}{Physics in Medicine and Biology} \textbf{\bibinfo{volume}{53}}, \bibinfo{pages}{1729} (\bibinfo{year}{2008}), ISSN \bibinfo{issn}{0031-9155, 1361-6560}, \urlprefix\url{https://iopscience.iop.org/article/10.1088/0031-9155/53/6/016}.

\bibitem[{\citenamefont{Geoghegan et~al.}(2020)\citenamefont{Geoghegan, Nelson, Flynn, Hill, Rana, and Hyer}}]{geoghegan_design_2020}
\bibinfo{author}{\bibfnamefont{T.~J.} \bibnamefont{Geoghegan}}, \bibinfo{author}{\bibfnamefont{N.~P.} \bibnamefont{Nelson}}, \bibinfo{author}{\bibfnamefont{R.~T.} \bibnamefont{Flynn}}, \bibinfo{author}{\bibfnamefont{P.~M.} \bibnamefont{Hill}}, \bibinfo{author}{\bibfnamefont{S.}~\bibnamefont{Rana}}, \bibnamefont{and} \bibinfo{author}{\bibfnamefont{D.~E.} \bibnamefont{Hyer}}, \bibinfo{journal}{Medical Physics} \textbf{\bibinfo{volume}{47}}, \bibinfo{pages}{2725} (\bibinfo{year}{2020}), ISSN \bibinfo{issn}{2473-4209}, \bibinfo{note}{\_eprint: https://onlinelibrary.wiley.com/doi/pdf/10.1002/mp.14139}, \urlprefix\url{https://onlinelibrary.wiley.com/doi/abs/10.1002/mp.14139}.

\bibitem[{\citenamefont{Nelson et~al.}(2023)\citenamefont{Nelson, Culberson, Hyer, Geoghegan, Patwardhan, Smith, Flynn, Yu, Gutiérrez, and Hill}}]{nelson_dosimetric_2023}
\bibinfo{author}{\bibfnamefont{N.~P.} \bibnamefont{Nelson}}, \bibinfo{author}{\bibfnamefont{W.~S.} \bibnamefont{Culberson}}, \bibinfo{author}{\bibfnamefont{D.~E.} \bibnamefont{Hyer}}, \bibinfo{author}{\bibfnamefont{T.~J.} \bibnamefont{Geoghegan}}, \bibinfo{author}{\bibfnamefont{K.~A.} \bibnamefont{Patwardhan}}, \bibinfo{author}{\bibfnamefont{B.~R.} \bibnamefont{Smith}}, \bibinfo{author}{\bibfnamefont{R.~T.} \bibnamefont{Flynn}}, \bibinfo{author}{\bibfnamefont{J.}~\bibnamefont{Yu}}, \bibinfo{author}{\bibfnamefont{A.~N.} \bibnamefont{Gutiérrez}}, \bibnamefont{and} \bibinfo{author}{\bibfnamefont{P.~M.} \bibnamefont{Hill}}, \bibinfo{journal}{Physics in Medicine \& Biology} \textbf{\bibinfo{volume}{68}}, \bibinfo{pages}{055003} (\bibinfo{year}{2023}), ISSN \bibinfo{issn}{0031-9155, 1361-6560}, \urlprefix\url{https://iopscience.iop.org/article/10.1088/1361-6560/acb6cd}.

\bibitem[{\citenamefont{Edwards and Syphers}(1993)}]{edwards_emittance_1993}
\bibinfo{author}{\bibfnamefont{D.~A.} \bibnamefont{Edwards}} \bibnamefont{and} \bibinfo{author}{\bibfnamefont{M.~J.} \bibnamefont{Syphers}}, in \emph{\bibinfo{booktitle}{An {Introduction} to the {Physics} of {High} {Energy} {Accelerators}}} (\bibinfo{publisher}{Wiley}, \bibinfo{year}{1993}), pp. \bibinfo{pages}{221--268}, ISBN \bibinfo{isbn}{978-3-527-61727-2}, \urlprefix\url{https://doi.org/10.1002/9783527617272.ch7}.

\bibitem[{\citenamefont{MacKinnon}(2009)}]{belsley_bootstrap_2009}
\bibinfo{author}{\bibfnamefont{J.~G.} \bibnamefont{MacKinnon}}, in \emph{\bibinfo{booktitle}{Handbook of {Computational} {Econometrics}}}, edited by \bibinfo{editor}{\bibfnamefont{D.~A.} \bibnamefont{Belsley}} \bibnamefont{and} \bibinfo{editor}{\bibfnamefont{E.~J.} \bibnamefont{Kontoghiorghes}} (\bibinfo{publisher}{John Wiley \& Sons, Ltd}, \bibinfo{address}{Chichester, UK}, \bibinfo{year}{2009}), pp. \bibinfo{pages}{183--213}, ISBN \bibinfo{isbn}{978-0-470-74891-6 978-0-470-74385-0}, \urlprefix\url{https://onlinelibrary.wiley.com/doi/10.1002/9780470748916.ch6}.

\bibitem[{\citenamefont{Steinberg et~al.}(2024)\citenamefont{Steinberg, Yap, Norman, Appleby, and Sheehy}}]{steinberg_characterising_2024}
\bibinfo{author}{\bibfnamefont{A.}~\bibnamefont{Steinberg}}, \bibinfo{author}{\bibfnamefont{J.}~\bibnamefont{Yap}}, \bibinfo{author}{\bibfnamefont{H.}~\bibnamefont{Norman}}, \bibinfo{author}{\bibfnamefont{R.}~\bibnamefont{Appleby}}, \bibnamefont{and} \bibinfo{author}{\bibfnamefont{S.}~\bibnamefont{Sheehy}}, \bibinfo{journal}{Nuclear Instruments and Methods in Physics Research Section A: Accelerators, Spectrometers, Detectors and Associated Equipment} \textbf{\bibinfo{volume}{1059}}, \bibinfo{pages}{169013} (\bibinfo{year}{2024}), ISSN \bibinfo{issn}{01689002}, \urlprefix\url{https://linkinghub.elsevier.com/retrieve/pii/S0168900223010136}.

\bibitem[{\citenamefont{{Vikas} and Sahu}(2021)}]{vikas_review_2021}
\bibinfo{author}{\bibnamefont{{Vikas}}} \bibnamefont{and} \bibinfo{author}{\bibfnamefont{R.~K.} \bibnamefont{Sahu}}, \bibinfo{journal}{Precision Engineering} \textbf{\bibinfo{volume}{71}}, \bibinfo{pages}{232} (\bibinfo{year}{2021}), ISSN \bibinfo{issn}{01416359}, \urlprefix\url{https://linkinghub.elsevier.com/retrieve/pii/S014163592100101X}.

\bibitem[{\citenamefont{Hübner and Wollnik}(1970)}]{hubner_design_1970}
\bibinfo{author}{\bibfnamefont{H.}~\bibnamefont{Hübner}} \bibnamefont{and} \bibinfo{author}{\bibfnamefont{H.}~\bibnamefont{Wollnik}}, \bibinfo{journal}{Nuclear Instruments and Methods} \textbf{\bibinfo{volume}{86}}, \bibinfo{pages}{141} (\bibinfo{year}{1970}), ISSN \bibinfo{issn}{0029554X}, \urlprefix\url{https://linkinghub.elsevier.com/retrieve/pii/0029554X70900455}.

\bibitem[{\citenamefont{Schermer}(1987)}]{schermer_tuning_1987}
\bibinfo{author}{\bibfnamefont{R.~I.} \bibnamefont{Schermer}}, \bibinfo{journal}{The 12th IEEE Particle Accelerator Conference,}  (\bibinfo{year}{1987}).

\bibitem[{\citenamefont{Caspi et~al.}(2014)\citenamefont{Caspi, Borgnolutti, Brouwer, Cheng, Dietderich, Felice, Godeke, Hafalia, Martchevskii, Prestemon et~al.}}]{caspi_cantedcosinetheta_2014}
\bibinfo{author}{\bibfnamefont{S.}~\bibnamefont{Caspi}}, \bibinfo{author}{\bibfnamefont{F.}~\bibnamefont{Borgnolutti}}, \bibinfo{author}{\bibfnamefont{L.}~\bibnamefont{Brouwer}}, \bibinfo{author}{\bibfnamefont{D.}~\bibnamefont{Cheng}}, \bibinfo{author}{\bibfnamefont{D.~R.} \bibnamefont{Dietderich}}, \bibinfo{author}{\bibfnamefont{H.}~\bibnamefont{Felice}}, \bibinfo{author}{\bibfnamefont{A.}~\bibnamefont{Godeke}}, \bibinfo{author}{\bibfnamefont{R.}~\bibnamefont{Hafalia}}, \bibinfo{author}{\bibfnamefont{M.}~\bibnamefont{Martchevskii}}, \bibinfo{author}{\bibfnamefont{S.}~\bibnamefont{Prestemon}}, \bibnamefont{et~al.}, \bibinfo{journal}{IEEE Transactions on Applied Superconductivity} \textbf{\bibinfo{volume}{24}}, \bibinfo{pages}{1} (\bibinfo{year}{2014}), ISSN \bibinfo{issn}{1558-2515}, \bibinfo{note}{conference Name: IEEE Transactions on Applied Superconductivity}, \urlprefix\url{https://ieeexplore.ieee.org/abstract/document/6621012}.

\bibitem[{\citenamefont{Wan et~al.}(2015)\citenamefont{Wan, Brouwer, Caspi, Prestemon, Gerbershagen, Schippers, and Robin}}]{wan_alternating-gradient_2015}
\bibinfo{author}{\bibfnamefont{W.}~\bibnamefont{Wan}}, \bibinfo{author}{\bibfnamefont{L.}~\bibnamefont{Brouwer}}, \bibinfo{author}{\bibfnamefont{S.}~\bibnamefont{Caspi}}, \bibinfo{author}{\bibfnamefont{S.}~\bibnamefont{Prestemon}}, \bibinfo{author}{\bibfnamefont{A.}~\bibnamefont{Gerbershagen}}, \bibinfo{author}{\bibfnamefont{J.~M.} \bibnamefont{Schippers}}, \bibnamefont{and} \bibinfo{author}{\bibfnamefont{D.}~\bibnamefont{Robin}}, \bibinfo{journal}{Physical Review Special Topics - Accelerators and Beams} \textbf{\bibinfo{volume}{18}}, \bibinfo{pages}{103501} (\bibinfo{year}{2015}), ISSN \bibinfo{issn}{1098-4402}, \urlprefix\url{https://link.aps.org/doi/10.1103/PhysRevSTAB.18.103501}.

\end{thebibliography}

\end{document}